\documentclass[twocolumn,10pt,a4paper,conference]{IEEEtran}
\usepackage[final]{microtype}
\usepackage[cmex10]{amsmath}
\usepackage{amssymb}   
\usepackage{amsxtra}
\usepackage{amscd}
\usepackage{amsthm}
\usepackage[acronym]{glossaries}
\usepackage{graphics}
\usepackage{graphicx}
\usepackage{epstopdf}
\usepackage{textcomp}
\usepackage{multirow}
\usepackage[export]{adjustbox}
\usepackage{bm}
\usepackage{lipsum}
\usepackage{url}
\usepackage{cite}
\usepackage{notoccite}
\usepackage{tikz}
\usetikzlibrary{shapes,backgrounds}
\usepackage{scalefnt}

\usepackage[font=footnotesize]{caption}
\usepackage[font=footnotesize]{subcaption}
\usepackage{dsfont}
\usepackage{enumitem}
\usepackage{tabularx}
\usepackage{mathtools}
\usepackage[normalem]{ulem}
\usepackage{MnSymbol}
\newacronym{IoT}{IoT}{internet-of-things}
\newacronym{D2D}{D2D}{device-to-device}
\newacronym{UMTS}{UMTS}{Universal Mobile Telecommunications System}
\newacronym{WLAN}{WLAN}{Wireless Local Area Network}
\newacronym{AP}{AP}{access point}
\newacronym{LiFi}{LiFi}{light-fidelity}
\newacronym{FET}{FET}{field-effect transistor}
\newacronym{PED}{PED}{passenger's electronic device}
\newacronym{MCRT}{MCRT}{Monte Carlo ray-tracing}
\newacronym{DL}{DL}{downlink}
\newacronym{UL}{UL}{uplink}
\newacronym{ICI}{ICI}{inter-channel interference}
\newacronym{LED}{LED}{light emitting diode}
\newacronym{PD}{PD}{photo-diode}
\newacronym{IFC}{IFC}{in-flight connectivity}
\newacronym{IFE}{IFE}{in-flight entertainment}
\newacronym{IM/DD}{IM/DD}{intensity-modulation and direct-detection}
\newacronym{IR}{IR}{infra-red}
\newacronym{VL}{VL}{visible light}
\newacronym{OW}{OW}{optical wireless}
\newacronym{OWC}{OWC}{optical wireless communication}
\newacronym{CAD}{CAD}{computer-aided-design}
\newacronym{VLC}{VLC}{visible light communications}
\newacronym{SIR}{SIR}{signal-to-interference-ratio}
\newacronym{SNR}{SNR}{signal-to-noise-ratio}
\newacronym{SNIR}{SNIR}{signal-to-noise-plus-interference-ratio}
\newacronym{QoS}{QoS}{quality-of-service}
\newacronym{RF}{RF}{radio frequency}
\newacronym{EM}{EM}{electromagnetic}
\newacronym{OOK}{OOK}{on-off keying}
\newacronym{PL}{PL}{path loss}
\newacronym{WDM}{WDM}{wavelength-division-multiplexing}
\newacronym{RGB}{RGB}{red-green-blue}
\newacronym{LoS}{LoS}{line-of-sight}
\newacronym{NLoS}{NLoS}{non-line-of-sight}
\newacronym{VoM}{VoM}{volume of mobility}
\newacronym{FoV}{FoV}{field-of-view}
\newacronym{SM}{SM}{spatial modulation}
\newacronym{GSM}{GSM}{generalized spatial modulation}
\newacronym{SSK}{SSK}{space shift keying}
\newacronym{GSSK}{GSSK}{generalized space shift keying}
\newacronym{SMX}{SMX}{spatial multiplexing}
\newacronym{OSM}{OSM}{optical spatial modulation}
\newacronym{MISO}{MISO}{multiple-input-single-output}
\newacronym{MIMO}{MIMO}{multiple-input-multiple-output}
\newacronym{OMIMO}{OMIMO}{optical MIMO}
\newacronym{VR}{VR}{virtual reality}
\newacronym{AR}{AR}{augmented reality}
\newacronym{MTC}{MTC}{machine type communications}
\newacronym{CAGR}{CAGR}{compound annual growth rate}
\newacronym{IM}{IM}{intensity modulation}
\newacronym{DD}{DD}{direct detection}
\newacronym{IMDD}{IM/DD}{intensity modulation and direct detection}
\newacronym{TX}{TX}{transmitter}
\newacronym{RX}{RX}{receiver}
\newacronym{PAM}{PAM}{pulse amplitude modulation}
\newacronym{MPAM}{$M$-PAM}{$M$-ary pulse amplitude modulation}
\newacronym{PAPR}{PAPR}{peak to average power ratio}
\newacronym{NRZ-OOK}{NRZ-OOK}{non-return-zero on-off keying}
\newacronym{O/E}{O/E}{optical-to-electrical conversion}
\newacronym{E/O}{E/O}{electrical-to-optical conversion}
\newacronym{DAC}{DAC}{digital-to-analog conversion}
\newacronym{ADC}{ADC}{analog-to-digital conversion}
\newacronym{DC}{DC}{direct-current}
\newacronym{SER}{SER}{symbol error ratio}
\newacronym{BER}{BER}{bit error ratio}
\newacronym{OFDM}{OFDM}{orthogonal frequency division multiplexing}
\newacronym{AWGN}{AWGN}{additive white Gaussian noise}
\newacronym{ML}{ML}{maximum-likelihood}
\newacronym{ZF}{ZF}{zero forcing}
\newacronym{MMSE}{MMSE}{minimum mean squared error}
\newacronym{ME-OSM}{ME-OSM}{minimum error OSM}
\newacronym{CCI}{CCI}{co-channel interference}
\newacronym{CCD}{CCD}{charge-coupled device}
\newacronym{UE}{UE}{user equipment}
\newacronym{PPM}{PPM}{pulse position modulation}
\newacronym{OA}{OA}{optical attocell}
\newacronym{ADR}{ADR}{angle diversity receiver}
\newacronym{FLIM}{FLIM}{flexible \gls{LED} index modulation}
\newacronym{GPSSK}{GPSSK}{generalized pulse position modulated \gls{SSK}}
\newacronym{GSM-MA}{GSM-MA}{\gls{GSM} multiple active}
\newacronym{PEP}{PEP}{pairwise error probability}
\newacronym{APEP}{APEP}{average pairwise error probability}
\newacronym{ABEP}{ABEP}{average bit error probability}
\newacronym{BEP}{BEP}{bit error probability}
\newacronym{FEC}{FEC}{forward error correction}
\newacronym{PMF}{PMF}{probability mass function}
\newacronym{PDF}{PDF}{probability density function}
\newacronym{CN}{CN}{condition number}
\newcommand\figref{Fig.~\ref}

\newcommand{\norm}[1]{ \vert\vert{#1} \rvert\rvert}

\newcommand{\sskicon}[1]{
	$\vcenter{\hbox{\begin{tikzpicture}
			\node[inner sep=0pt] (venn) at (0,0)
			{\includegraphics[#1]{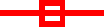}};
			\end{tikzpicture}}}$}

\newcommand{\gsskiicon}[1]{
	$\vcenter{\hbox{\begin{tikzpicture}
			\node[inner sep=0pt] (venn) at (0,0)
			{\includegraphics[#1]{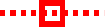}};
			\end{tikzpicture}}}$}

\newcommand{\gsskiiicon}[1]{
	$\vcenter{\hbox{\begin{tikzpicture}
			\node[inner sep=0pt] (venn) at (0,0)
			{\includegraphics[#1]{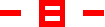}};
			\end{tikzpicture}}}$}

\newcommand{\smicon}[1]{
	$\vcenter{\hbox{\begin{tikzpicture}
			\node[inner sep=0pt] (venn) at (0,0)
			{\includegraphics[#1]{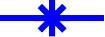}};
			\end{tikzpicture}}}$}

\newcommand{\gsmicon}[1]{
	$\vcenter{\hbox{\begin{tikzpicture}
			\node[inner sep=0pt] (venn) at (0,0)
			{\includegraphics[#1]{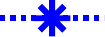}};
			\end{tikzpicture}}}$
}

\newcommand{\gsmiiicon}[1]{
	$\vcenter{\hbox{\begin{tikzpicture}
			\node[inner sep=0pt] (venn) at (0,0)
			{\includegraphics[#1]{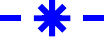}};
			\end{tikzpicture}}}$}

\newcommand{\smxicon}[1]{
	$\vcenter{\hbox{\begin{tikzpicture}
			\node[inner sep=0pt] (venn) at (0,0)
			{\includegraphics[#1]{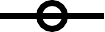}};
			\end{tikzpicture}}}$}

\newcommand{\flimicon}[1]{
	$\vcenter{\hbox{\begin{tikzpicture}
			\node[inner sep=0pt] (venn) at (0,0)
			{\includegraphics[#1]{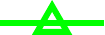}};
			\end{tikzpicture}}}$}

\newcounter{tempEquationCounter} 
\newcounter{thisEquationNumber}

\IEEEoverridecommandlockouts

\begin{document}
%
\title{Flexible LED Index Modulation for MIMO Optical Wireless Communications}
%
%
%

\author{\IEEEauthorblockN{Anil~Yesilkaya\IEEEauthorrefmark{1}, Ardimas~Andi~Purwita\IEEEauthorrefmark{1}, Erdal~Panayirci\IEEEauthorrefmark{2}, H. Vincent~Poor\IEEEauthorrefmark{3}~and~Harald~Haas\IEEEauthorrefmark{1}}
	\IEEEauthorblockA{\IEEEauthorrefmark{1}Institute for Digital Communications, LiFi R\&D Centre, The University of Edinburgh, Edinburgh EH9 3JL, UK 
	\\ Email: \{a.yesilkaya, a.purwita, h.haas\}@ed.ac.uk}
	\IEEEauthorblockA{\IEEEauthorrefmark{2}Department of Electrical and Electronics Engineering, Kadir Has University, 34083, Istanbul, Turkey
	\\ Email: eepanay@khas.edu.tr}
	\IEEEauthorblockA{\IEEEauthorrefmark{3}Department of Electrical Engineering, Princeton University, NJ-08544, USA 
	\\ Email: poor@princeton.edu }
	\thanks{This work was supported by EPSRC under Established Career Fellowship Grant \mbox{EP/R007101/1}. This work was also supported in part by the Scientific and Technical Research Council of Turkey (TUBITAK) under the 1003-Priority Areas R\&D Projects support Program No. \mbox{218E034} and KAUST under Grant No. \mbox{OSR-2016-CRG5-2958-02}. A. Yesilkaya acknowledges the financial support from Zodiac Inflight Innovations (TriaGnoSys GmbH). \mbox{A. A. Purwita} acknowledges the financial support from Indonesian Endowment Fund for Education (LPDP). H. Haas acknowledges support from the Wolfson Foundation and the Royal Society.}
}

%
%

\markboth{Journal of \LaTeX\ Class Files,~Vol.~14, No.~8, August~2015}%
{Shell \MakeLowercase{\textit{et al.}}: Bare Demo of IEEEtran.cls for IEEE Journals}
%


\IEEEoverridecommandlockouts
\IEEEpubid{\makebox[\columnwidth]{\copyright~2021 IEEE. Digital Object Identifier 10.1109/GLOBECOM42002.2020.9322528 \hfill} \hspace{\columnsep}\makebox[\columnwidth]{ }}

\maketitle

\begin{abstract}
	The limited bandwidth of \gls{OWC} front-end devices motivates the use of \gls{MIMO} techniques to enhance data rates. It is known that very high multiplexing gains can be achieved by \gls{SMX} at the cost of prohibitive detection complexity. Alternatively, in \gls{SM}, a single \gls{LED} is activated per time instance where information is carried by both the signal and the \gls{LED} index. Since only one \gls{LED} is active, both the \gls{TX} and \gls{RX} complexity reduce significantly while retaining the information transmission in the spatial domain. However, this simplified \gls{TX} utilization approach leads \gls{SM} to suffer from significant spectral efficiency losses compared to \gls{SMX}. In this paper, we propose a technique that benefits from the advantages of both systems. Accordingly, the proposed \gls{FLIM} technique harnesses the inactive state of the \glspl{LED} as a transmit symbol. Therefore, the number of active \glspl{LED} changes in each transmission, unlike conventional techniques. Moreover, the system complexity is reduced by employing a linear \gls{MMSE} equalizer and an angle perturbed receiver. Numerical results show that \gls{FLIM} outperforms the reference systems by at least $6$ dB in the low and medium/high spectral efficiency regions.
\end{abstract}

\begin{IEEEkeywords}
	Optical wireless communications (OWC), multiple-input-multiple-output (MIMO), spatial modulation (SM), minimum mean square equalizer (MMSE).
\end{IEEEkeywords}

%
\IEEEpeerreviewmaketitle
\glsresetall
\section{Introduction}
\Gls{SM} is a \gls{MIMO} transmission technique which avoids \gls{ICI} and achieves better power efficiency and error performance \cite{4149911}. In conventional \gls{SM}, all but one of the \gls{TX} units are silenced at a given transmit symbol. Hence, the \gls{ICI} is completely avoided while sending information by both the transmitted signal (\textit{constellation symbols}) and the active \gls{TX} index (\textit{spatial symbols}) simultaneously. Moreover, the detection complexity at the \gls{RX} is significantly reduced in \gls{SM} as each symbol is transmitted by a single transmit unit. The complexity of \gls{SM} could be further decreased by omitting the constellation symbols, which leads to \gls{SSK} \cite{956483}. In \gls{SSK}, the detection complexity at the \gls{RX} is simplified in exchange for reduced spectral efficiency. In \cite{5165332}, it is reported that the error performance of \gls{SM} and \gls{SSK} are almost equal, which makes \gls{SSK} a good candidate for low/medium data rate applications. In order to increase the spectral efficiency of \gls{SSK}, \gls{GSSK} is proposed in \cite{4699782} where multiple \glspl{TX} are activated per transmission instance. Similarly, generalization of \gls{SM}, having multiple active transmit units is also proposed in \cite{5757786} and \cite{5700442} independently. Note that in \gls{GSM}, all the active \glspl{TX} send the same signal which avoids \gls{ICI}. The spectral efficiency of \gls{GSM} is further enhanced by choosing different constellation symbols for each active \gls{TX} in \cite{6166339,6831300}. Two independent proposals of this idea, namely \gls{GSM} with multiple active antennas in \cite{6166339} and multi-stream SM in \cite{6831300}, are referred to as \gls{GSM}-II throughout this paper. Also, the spectral efficiency of \gls{GSSK} is improved in optical communications for a $2\times 2$ system in \cite{6476612}\footnote{This is the more generalized version of \gls{GSSK} and will be referred to as \gls{GSSK}-II throughout the paper. Note that the maximum spectral efficiency case for \gls{GSSK}-II, when the duty cycle $\tau=1$ \cite{6292784}, is considered in this paper.}. Accordingly, an orthogonal transmission symbol set is proposed to increase the error performance by combining \gls{SSK} and \gls{PPM}. It is shown that \gls{GSSK}-II can achieve a linear spectral efficiency with respect to the number of transmit units, whereas all the other mentioned systems so far can only achieve a logarithmic spectral efficiency gain with increased number of \glspl{TX}. However, the generalization of \gls{GSSK}-II to any number of transmit units is not straightforward and the orthogonality of the signal set could only be provided for higher order modulations of \gls{PPM}. Hence, the error performance of the system will decrease. The constellation design techniques for $2\times 2$ and $3\times3$ \gls{MIMO}-\gls{OWC} are given in \cite{7812635} and \cite{ZHU2017260}, respectively. However, the proposed technique requires a careful design procedure where an arbitrary number of \glspl{LED} and constellation orders are not supported. Furthermore, the effect of user mobility on the system performance is not considered.

In this paper, a \gls{MIMO} transmission method for \gls{OWC}, referred to as \gls{FLIM}, is proposed. In \gls{FLIM}, the number of active \glspl{TX} is not fixed, which greatly extends the transmission possibilities. Moreover, \gls{FLIM} brings the flexibility to adjust power and/or error performance simply by choosing a proper transmission signal subset. In order to have a feasible detection complexity, a linear \gls{MMSE} detector is used at the receiver. Lastly, a perturbed receiver ensures that the channel matrix of a mobile user is non-singular.

The rest of the paper is structured as follows: In Section II, the \gls{MIMO} optical channel structure is described. The details of \gls{SM}-based \gls{MIMO} systems, including the proposed \gls{FLIM}, is given in Section III. The \gls{ABEP} performance of \gls{FLIM} is compared with reference systems in Section IV. Conclusions are drawn in Section V.

\emph{Notation}: Matrices and column vectors are written in bold uppercase and bold lowercase letters, respectively. The $m^\textrm{th}$ row and $n^\textrm{th}$ column element of a matrix $\mathbf{A}$ is denoted by $A_{m,n}$. Similarly, the $m^\textrm{th}$ element of a vector $\mathbf{a}$ is given as $a_m$. The trace of a matrix, transpose of a matrix/vector and the Euclidean norm of a vector are expressed as $\text{tr}(\cdot)$, $(\cdot)^\textrm{T}$ and $\norm{\cdot}$, respectively. A real normal distribution with mean $\mu$ and variance $\sigma^2$ is denoted by $\mathcal{N}(\mu,\sigma^2)$. The statistical expectation, argument maximum, argument minimum, flooring, dot product operators, Q-function, and the set of real numbers are given by $\textrm{E}\{\cdot\}$, $\arg\max\{\cdot\}$, $\arg\min\{\cdot\}$, $\lfloor \cdot \rfloor$, $\boldsymbol{\cdot}$, $Q(\cdot)$, and $\mathbb{R}$, respectively. Lastly, the $m\times m$ identity matrix and $n\times 1$ all-zeroes vector are denoted by $\mathbf{I}_{m}$ and $\mathbf{0}_n$, respectively.
\section{MIMO-OWC Channel}
\begin{figure}[!t]
	\centering
	\begin{subfigure}[b]{0.95\columnwidth}
		\centering
		\includegraphics[width=1\columnwidth,draft=false]{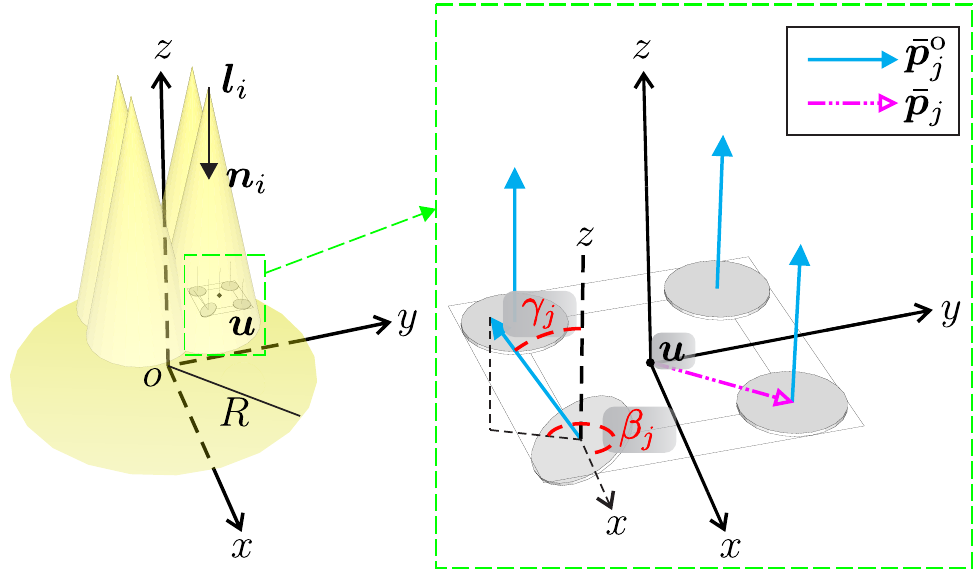}
		\caption{Geometric models of adopted LEDs and PDs}
	\end{subfigure}\\
	\begin{subfigure}[b]{0.95\columnwidth}
		\centering
		\includegraphics[width=1\columnwidth,draft=false]{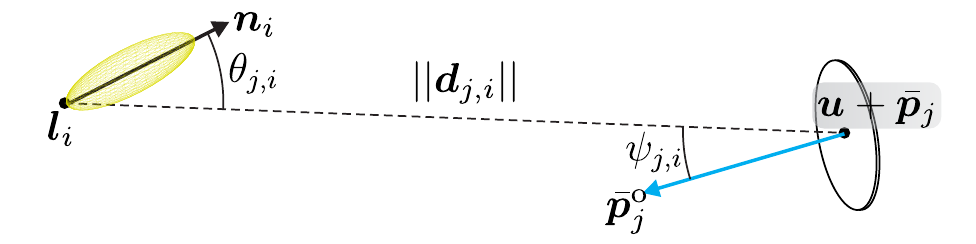}
		\caption{Link geometry}
	\end{subfigure}
	\caption{Geometric details of the MIMO-OWC channel.}
	\label{fig_ledpdmodel}
\end{figure}
In this work, we consider an optical \gls{MIMO} application where each mobile \gls{UE} is equipped with non-imaging receivers. In \figref{fig_ledpdmodel}(a), the system structure is depicted where the luminaire is serving as the optical \gls{AP} and the \gls{UE} can be in any location within the attocell boundaries. Each optical \gls{AP} and mobile \gls{UE} is assumed to be equipped with $N_\textrm{t}$ \glspl{LED} and $N_\textrm{r}$ \glspl{PD}, respectively. As reported in \cite{7362097}, the \gls{LoS} link is sufficient to accurately model a practical \gls{OWC} system. Accordingly, the multipath signal contributions are negligible as long as the \gls{UE} is not located at the corners of the room. The geometric parameters of the \glspl{LED} and \glspl{PD} are given in vector notation where the origin is depicted in \figref{fig_ledpdmodel}(a) as the point $o$. The unit normal vectors in $x$, $y$ and $z$ axes are given as $\bm{n}_x$, $\bm{n}_y$ and $\bm{n}_z$, respectively. The locations of the $i^{\textrm{th}}$ \gls{LED}, $j^{\textrm{th}}$ \gls{PD} and \gls{UE} are denoted by their position vectors as $\bm{l}_{i}$, $\bm{p}_{j}$ and $\bm{u}$, respectively. Moreover, the unit normal vector of the $i^\textrm{th}$ \gls{LED} is given by $\bm{n}_i$. All the \glspl{PD} are assumed to be placed on the same plane as depicted in \figref{fig_ledpdmodel}(a). The relationship between $\bm{u}$ and $\bm{p}_{j}$ is given by $\bm{p}_j=\bm{u}+\bar{\bm{p}}_j$, where $\bar{\bm{p}}_j$ is the position vector of the $j^{\textrm{th}}$ \gls{PD} relative to the \gls{UE}\footnote{The origin for the vectors with bar on top is the point $\bm{u}$.}. The orientation of the $j^{\textrm{th}}$ \gls{PD} is defined by the unit orientation vector $\bar{\bm{p}}^{\textrm{o}}_j$. The polar angle of the $j^{\textrm{th}}$ \gls{PD} with respect to $\mathbf{n}_z$ is defined by $\gamma_j=\bm{n}_z \boldsymbol{\cdot} \bar{\bm{p}}^{\textrm{o}}_j$. Similarly, the azimuth angle of the $j^{\textrm{th}}$ \gls{PD} with respect to $\bm{n}_x$ is given as $\beta_j=\mathbf{n}_x \boldsymbol{\cdot} \tilde{\bm{p}}^{\textrm{o}}_j$ where $\tilde{\bm{p}}^{\textrm{o}}_j$ is the projection of the $\bar{\bm{p}}^{\textrm{o}}_j$ onto the $xy$-plane. Thus, the orientation vector could be expressed as $\bar{\bm{p}}^\textrm{o}_j=\left[\sin(\gamma_j)\cos(\beta_j),~\sin(\gamma_j)\sin(\beta_j),~\cos(\gamma_j) \right]$. It should be noted that all the \glspl{LED} in a luminaire are assumed to be facing downward ($-\bm{n}_z$ direction). Also, both the \glspl{LED} and \glspl{PD} are assumed to be located at the corners of a square plane, which is the most common case in practical applications. Therefore, the \gls{LoS} \gls{DC} channel gain between the $i^{\textrm{th}}$ \gls{LED} and the $j^{\textrm{th}}$ \gls{PD} is given as follows \cite{554222,ybth1901}:
\begin{align}\label{eq_dcchannelgain}
H_{j,i} = \frac{(m+1)A_{\textrm{PD}}}{2 \pi \norm{ \bm{d}_{j,i} }^{m+3} }\left( -\bm{n}_{z}\boldsymbol{\cdot} \bm{d}_{j,i} \right)^m \left( \bar{\bm{p}}^\textrm{o}_j \boldsymbol{\cdot}\bm{d}_{j,i} \right) \mathds{1}_{\kappa_{1/2}}\left(\psi_{j,i} \right),
\end{align}
\noindent where $i\in \{1,2,\cdots,N_\textrm{t}\}$ and $j\in\{1,2,\cdots,N_\textrm{r}\}$. The Lambertian mode number is defined as, $m=-1/\log_2(\Phi_{1/2})$ where $\Phi_{1/2}$ is the semi-angle of half power of the \gls{LED}. The detector area of the \glspl{PD} are denoted by $A_{\textrm{PD}}$. The angle of incidence between the $i^\text{th}$ \gls{LED} and the $j^\text{th}$ \gls{PD}, depicted in \figref{fig_ledpdmodel}(b), is denoted as $\psi_{j,i}$. The Euclidean distance vector directed from the $j^\text{th}$ \gls{PD} to the $i^\text{th}$ \gls{LED} is given as, $\bm{d}_{j,i} =\bm{l}_i-\bm{p}_j=\bm{l}_i-\bm{u}-\bar{\bm{p}}_j$. The indicator function, $\mathds{1}_{\kappa_{1/2}}\left(\cdot \right)$, defines the \gls{FoV} of \glspl{PD} as follows:
\[
\mathds{1}_{\kappa_{1/2}}\left(x \right) = \left\{\begin{array}{lr}
1, & \text{if } \vert x \vert\leq \kappa_{1/2}\\
0, & \text{otherwise}
\end{array}\right.
\]
\noindent where $\kappa_{1/2}$ is the half angle of the \gls{FoV} of the \glspl{PD}.
\noindent Consequently, the optical channel matrix becomes
\begin{equation}
\mathbf{H} =
\begin{bmatrix}
H_{1,1} & \dots & H_{1,N_{\text{t}}}\\
\vdots & \ddots & \vdots\\
H_{N_{\text{r}},1} & \dots & H_{N_{\text{r}},N_{\text{t}}}
\end{bmatrix},
\label{eqchannelmatrix}
\end{equation}
\noindent where $0\leq H_{j,i}\leq 1$ and $H_{j,i} \in \mathbb{R}^+, ~~\forall i,j$.
\section{System Description}
\begin{figure}[!t]
	\centering
	\begin{subfigure}[b]{.5\columnwidth}
		\centering
		\includegraphics[width=0.9\columnwidth,draft=false]{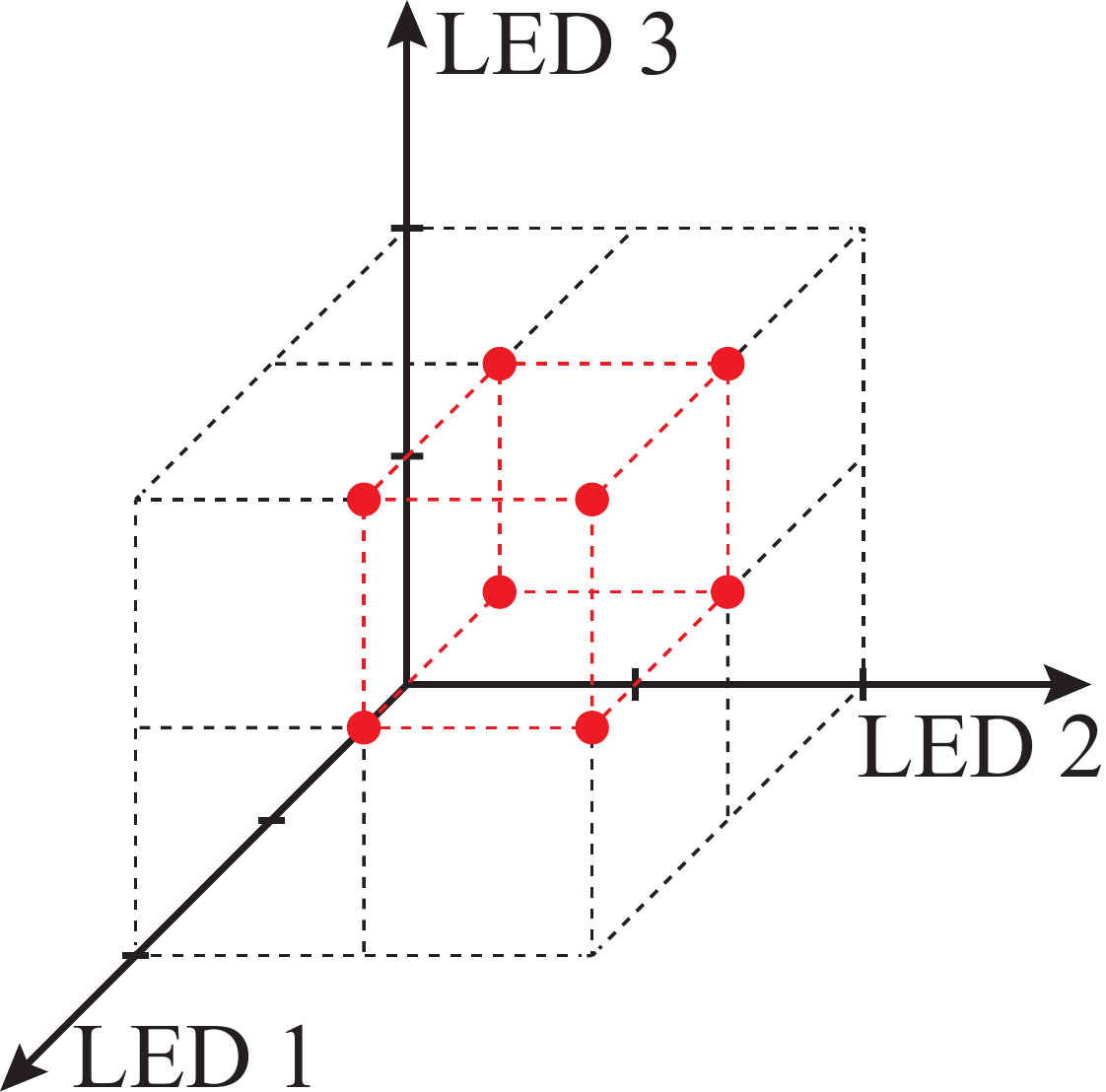}
		\caption{SMX}
	\end{subfigure}~
	\begin{subfigure}[b]{.5\columnwidth}
		\centering
		\includegraphics[width=0.9\columnwidth,draft=false]{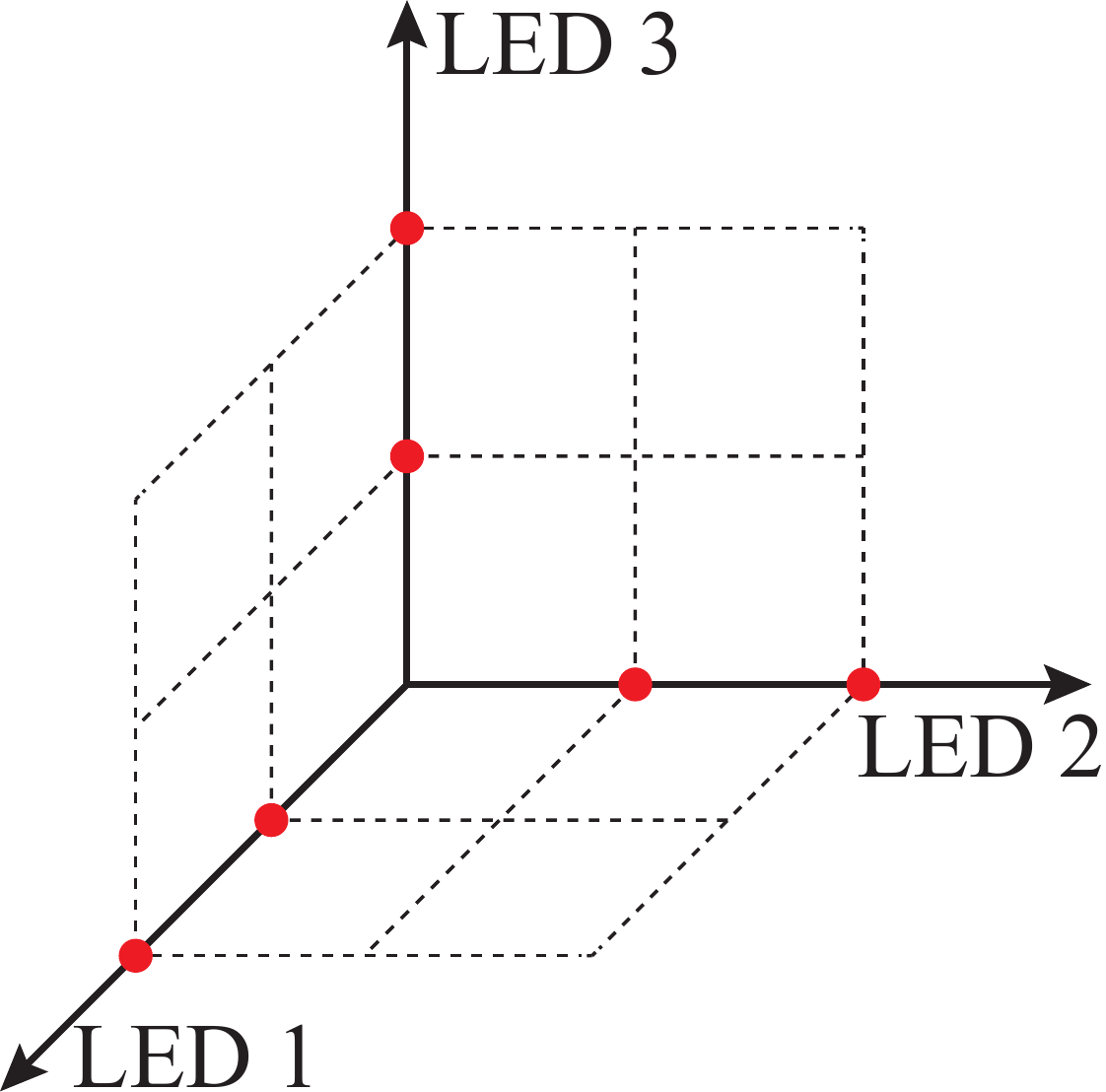}
		\caption{SM}
	\end{subfigure}\\ \vspace{4mm}
	\begin{subfigure}[b]{.5\columnwidth}
		\centering
		\includegraphics[width=0.9\columnwidth,draft=false]{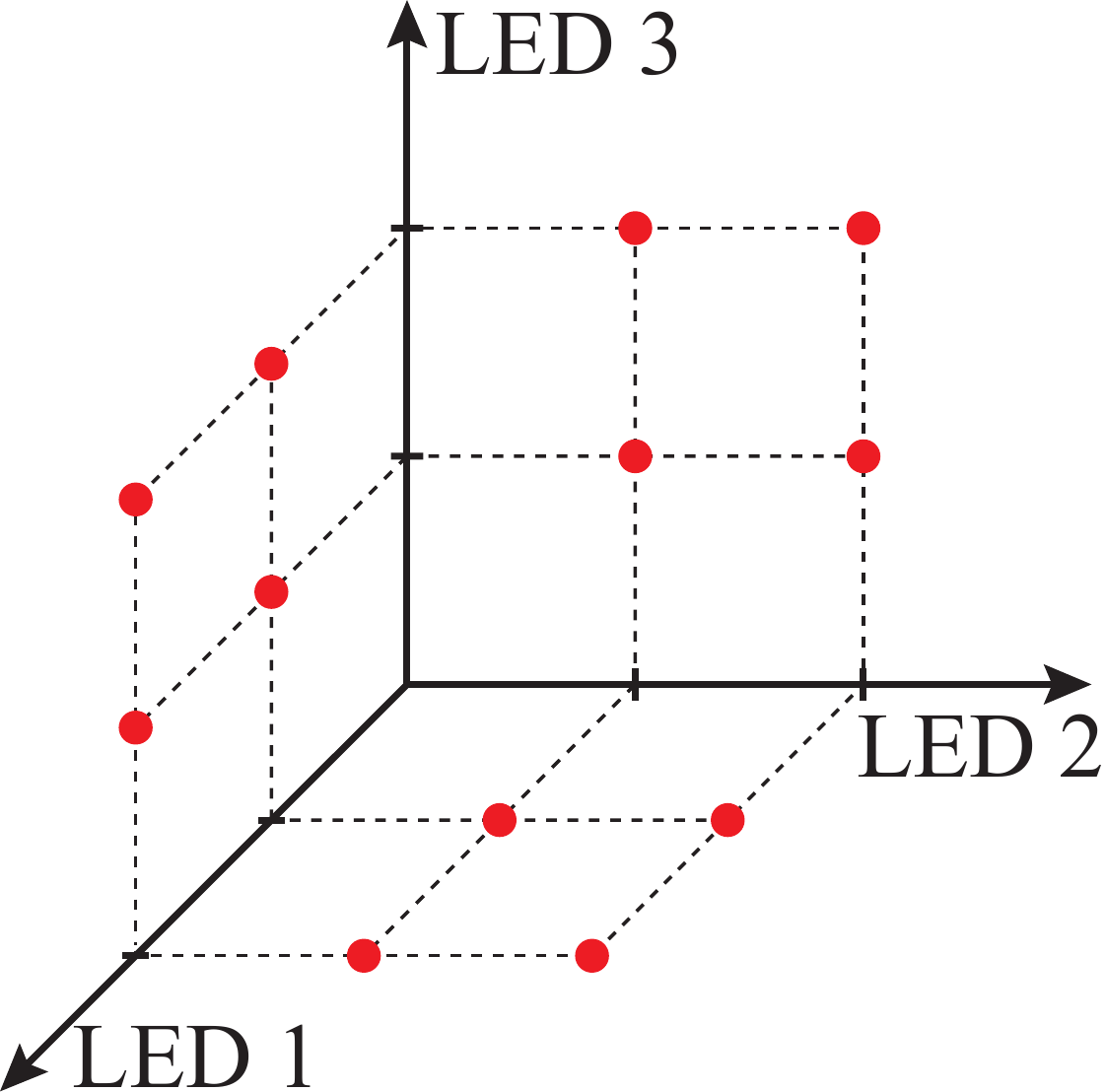}
		\caption{GSM-II}
	\end{subfigure}~
	\begin{subfigure}[b]{.5\columnwidth}
		\centering
		\includegraphics[width=0.9\columnwidth,draft=false]{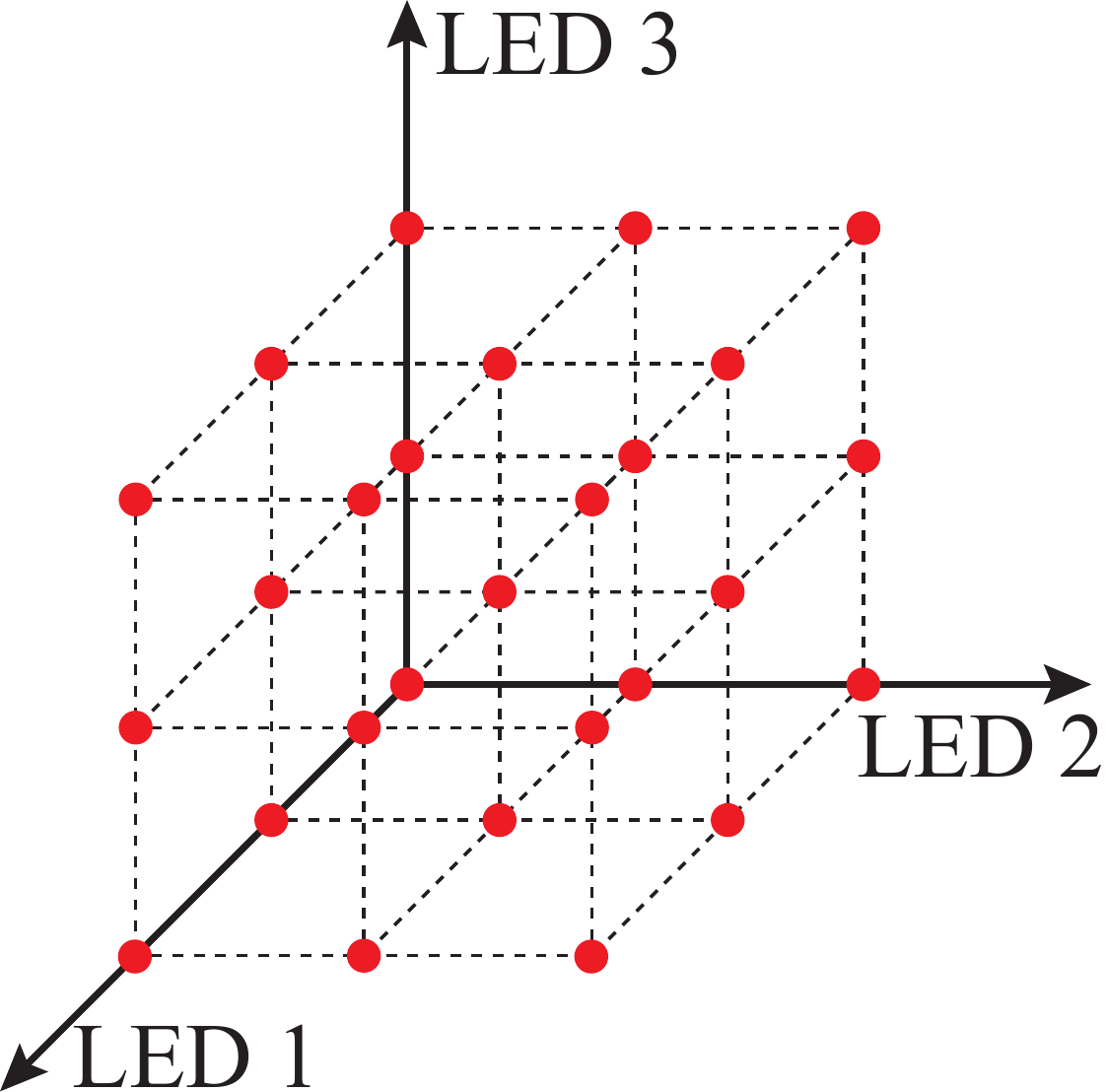}
		\caption{FLIM}
	\end{subfigure}
	\caption{Transmission symbol possibilities are depicted for (a) SMX, (b) SM, (c) GSM-II and (d) FLIM. The red dots are depicting the possible PAM symbols could be transmitted from the LEDs when $M=2$ and $N_\text{t}=3$.}
	\label{fig_signalspace}
\end{figure}

The $N_\textrm{t}\text{-LED}\times N_\textrm{r}\text{-PD}$ \gls{FLIM} technique is detailed in this section. Accordingly, $N_\textrm{b}$-bits binary information vector $\mathbf{b}_u = \left[ b_{u,0},b_{u,1},\cdots,b_{u,N_\textrm{b}-1} \right]^\text{T}$ is generated by the user and fed into a mapping function $f_\mathcal{M}$. The \textit{one-to-one} mapping rule encodes each $\mathbf{b}_u$ onto a $N_\textrm{t}$-length transmission vector $\mathbf{s}_t = \left[ s_{t,0},s_{t,1},\cdots,s_{t,N_\textrm{t}-1} \right]^\text{T}$, where $\mathbf{b}_u\in \mathbb{B}$, $\mathbf{s}_t \in \mathbb{S}$ for $1\leq u,t \leq C$. The binary permutation vector set and transmit vector set are denoted by $\mathbb{B}$ and $\mathbb{S}$, respectively. The cardinality of the sets are determined by the chosen modulation technique as $\vert \mathbb{B} \vert=\vert \mathbb{S} \vert=C$. It should be noted that $C=2^n$ where $n\in \mathbb{Z}$. In some modulations such as \gls{GSM}, \gls{GSM}-II, \gls{GSSK} and \gls{FLIM}, the total number of symbols, $K$, is not a power of two. Hence, a subset of the transmit symbols must be chosen such that $C=2^{\lfloor \log_2(K) \rfloor}$. The number of bits per transmit symbol becomes $N_\textrm{b}=\log_2(C)$. The mapping function $f_\mathcal{M}$ is given by $f_\mathcal{M}:\mathbf{b}_u \in \mathbb{B} \rightarrow \mathbf{s}_t \in \mathbb{S}$, where $b_{u,i}\in \{0,1\}$ and $s_{t,j}\in \mathcal{M}$ for $0\leq i \leq N_\textrm{b}-1$, \mbox{$0 \leq j \leq N_\textrm{t}-1$}. It is important to note that $\mathbf{s}_t$ conveys both the constellation and spatial symbols also, $\mathcal{M}$ denotes the finite modulation alphabet. As the \gls{IMDD} signals are constrained to be real and positive valued, the forward driving current of the \glspl{LED}, $I_\textrm{f}$, is assumed to be modulated by unipolar $M$-ary \gls{PAM}. Thus, the unipolar $M$-\gls{PAM} alphabet is given by
\begin{align}
\mathcal{M} = \left\{I_\textrm{f}\in \mathbb{R}^+ : I_{\textrm{L}} + \frac{ I_{\textrm{U}}-I_{\textrm{L}} }{ M-1 } k,\quad  k \in \{0,1,\cdots, M-1\} \right\},
\label{eq_pam}
\end{align}
\noindent where $M$ is the \gls{PAM} order. The upper and lower limits of the forward driving current are given by $I_{\textrm{U}}$ and $I_{\textrm{L}}$, respectively. It is reported in \cite{Qian:15} that the error performance of \gls{PAM} is highly dependent on the non-linear characteristics of the \glspl{LED}. Therefore, it is important to choose $I_{\textrm{U}}$ and $I_{\textrm{L}}$ within the dynamic range of an \gls{LED}. In \gls{SM}-based systems, $I_\textrm{f}$ could be zero which corresponds to the \textit{inactive} state of an \gls{LED}. The locations of the \textit{inactive} \glspl{LED} and values of the \textit{active} \glspl{LED} are the unique signature of the adopted modulation technique. Thus, the transmission vector set for \gls{SM} is given for $N_\textrm{t}=3$ as follows:
\begin{align*}
\mathbb{S}^\text{SM} = \left\{ 
\begin{bmatrix}
m_0 \\
0 \\
0
\end{bmatrix},
\begin{bmatrix}
0 \\
m_1 \\
0
\end{bmatrix},
\begin{bmatrix}
0 \\
0 \\
m_2
\end{bmatrix}
\right\}.
\end{align*}
\noindent The transmitted constellation symbols are given by \mbox{$m_i\in \mathcal{M},~~\forall i$}. Note that the $N_\text{t}=3$ is only chosen for the presentation simplicity, refer to Fig. \ref{fig_signalspace}, in practice $N_\text{t}$ is power of two to achieve the adequate bit labelling. Similar to \gls{SM}, the transmission vector sets for \gls{GSM}-II and \gls{SMX} are given by
\begin{align*}
\mathbb{S}^\text{GSM-II}(2) = \left\{ 
\begin{bmatrix}
m_0 \\
m_1 \\
0
\end{bmatrix},
\begin{bmatrix}
m_2 \\
0 \\
m_3
\end{bmatrix},
\begin{bmatrix}
0 \\
m_4 \\
m_5
\end{bmatrix}
\right\}\quad\textrm{and}\quad
\mathbb{S}^\text{SMX} =
\begin{bmatrix}
m_0 \\
m_1 \\
m_2
\end{bmatrix},
\end{align*}
\noindent
where the \gls{GSM}-II transmit vector for $N_\textrm{a}$ active transmitters is denoted as $\mathbb{S}^\text{GSM-II}(N_\textrm{a})$. It should be noted that both \gls{SM} and \gls{SMX} are the special cases of \gls{GSM}-II with \mbox{$\mathbb{S}^\text{SM}=\mathbb{S}^\text{GSM-II}(1)$} and \mbox{$\mathbb{S}^\text{SMX}=\mathbb{S}^\text{GSM-II}(N_\textrm{t})$}. In \gls{FLIM}, the constraint of a predetermined number of active \glspl{LED} is relaxed, $N_\text{a}\in \{0,1,2,\cdots,N_\text{t}\}$, which extends the transmission alphabet significantly. Thus, the transmission vector set for \gls{FLIM} becomes
\begin{align}
\mathbb{S}^\text{FLIM} = \left\{ \mathbf{0}_{N_\textrm{t}}, \mathbb{S}^\text{SM}, \mathbb{E}^{\text{GSM-II}}, \mathbb{S}^\text{SMX}  \right\},
\label{eq_setFLIM}
\end{align}
\noindent where the universal set for the activation patterns of \gls{GSM}-II is given by:
\begin{align}
\mathbb{E}^{\text{GSM-II}}=\bigcup\limits_{N_\textrm{a}=2}^{N_{\textrm{t}}-1}\mathbb{S}^\text{GSM-II}(N_\textrm{a}).
\end{align}
\noindent It should be noted that all-off, $\mathbf{0}_{N_\textrm{t}}$, and the all-on, $\mathbb{S}^\text{SMX}$, patterns are strictly avoided in \gls{RF} based \gls{SM} systems \cite{6428727,7867774}. However, the information carrying signal can be zero or near zero in \gls{OWC} due to the \gls{LED} turn-on current. The all-on pattern is also utilized in \gls{FLIM} in order to enlarge the transmit vector set. The larger transmission vector set brings the advantage of flexibility to \gls{FLIM} by choosing a subset to optimize: either error performance or power efficiency. The transmit symbol spaces for the aforementioned \gls{SM}-based systems including \gls{FLIM} is depicted for the cases where $N_\textrm{t}=3$ and $M=2$ in \figref{fig_signalspace}. It can be seen from the figure that \gls{FLIM} contains the symbols of all the other systems along with its unique symbols. Therefore, the spectral efficiency of \gls{FLIM} becomes
\begin{align}
\eta_\textrm{FLIM}=\lfloor N_\textrm{t}\log_2(M+1) \rfloor \quad \textrm{bits per channel use (bpcu)}.
\label{eq_etaflim}
\end{align}
\begin{figure}[!t]
	\centering
	\begin{subfigure}[b]{1\columnwidth}
	\centering
	\includegraphics[width=1\columnwidth,draft=false]{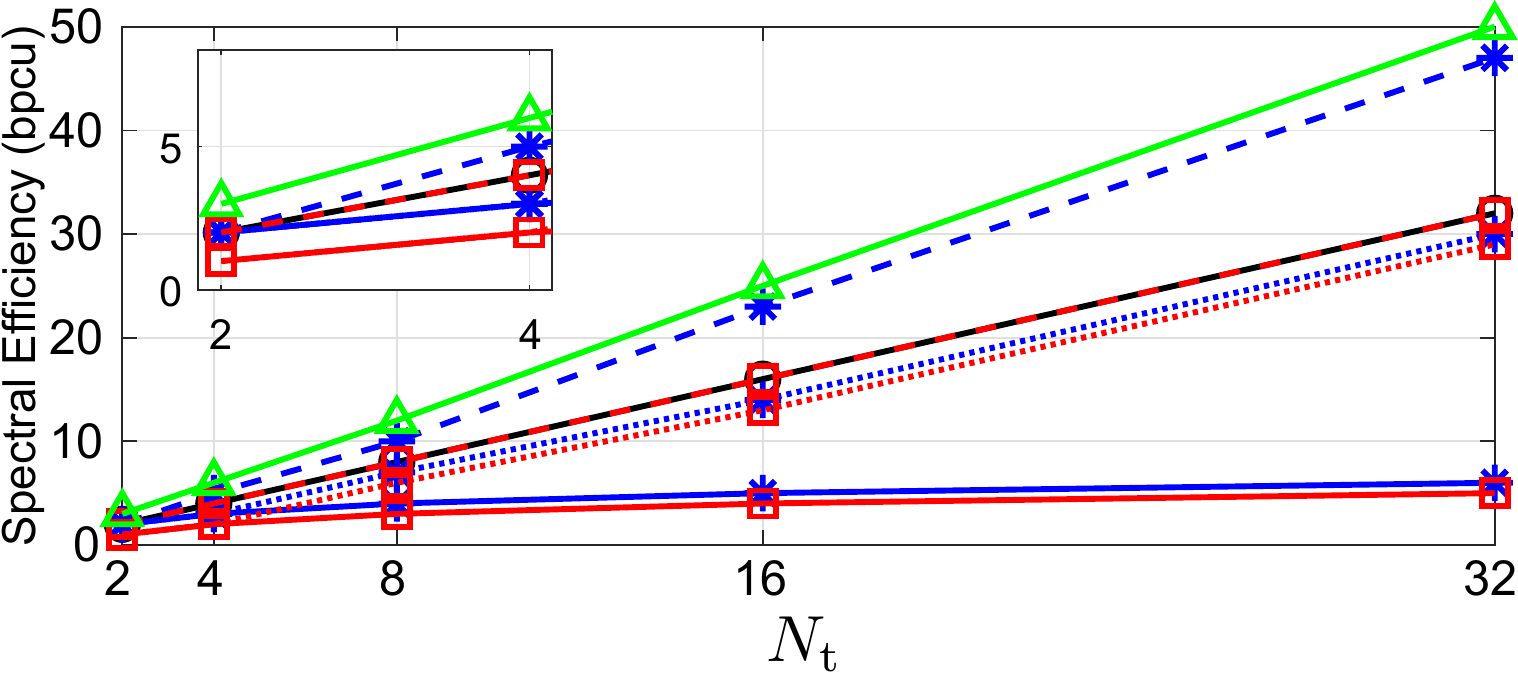}
	\caption{$M=2$}
	\end{subfigure}\\
	\begin{subfigure}[b]{1\columnwidth}
	\centering
	\includegraphics[width=1\columnwidth,draft=false]{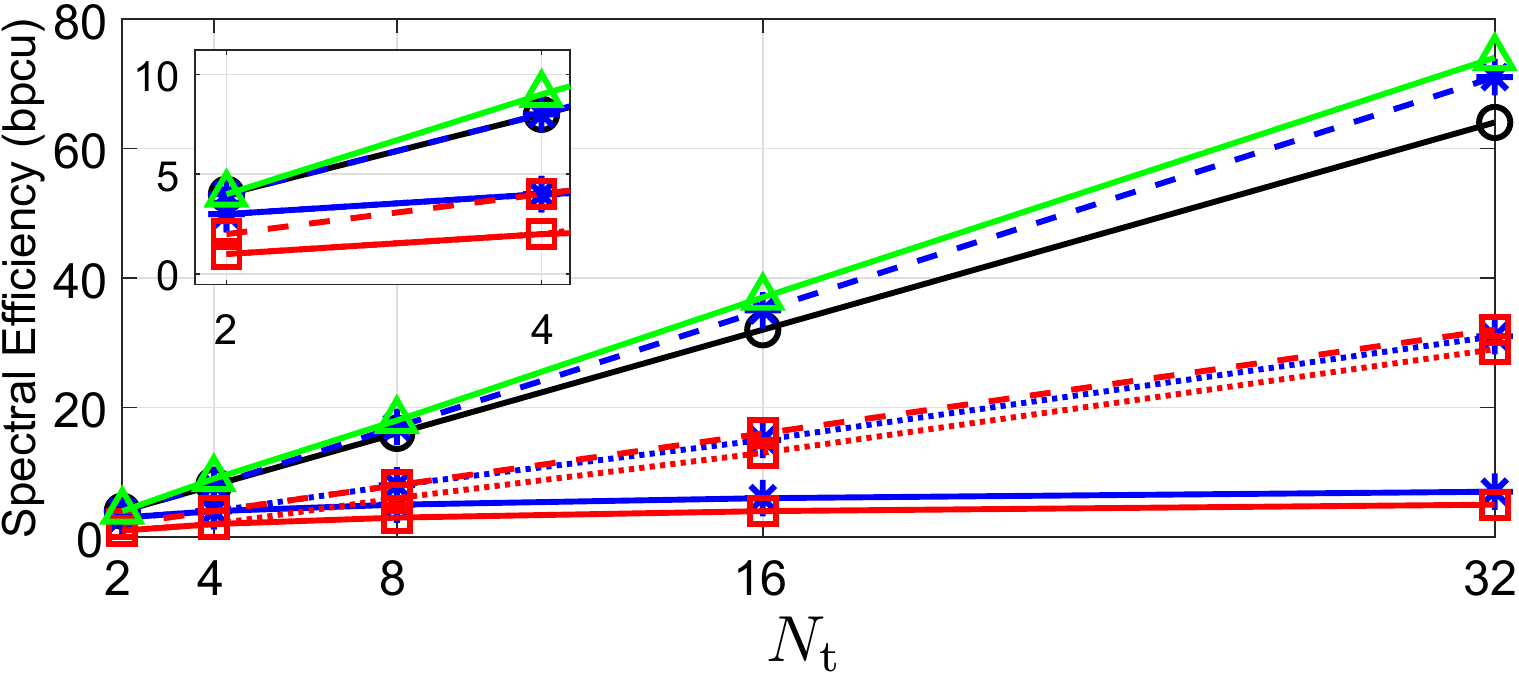}
	\caption{$M=4$}
	\end{subfigure}
	\caption{Spectral efficiency comparison for SSK (\protect\sskicon{scale=0.7}), GSSK (\protect\gsskiicon{scale=0.7}), GSSK-II (\protect\gsskiiicon{scale=0.7}), SM (\protect\smicon{scale=0.7}), GSM (\protect\gsmicon{scale=0.7}), GSM-II (\protect\gsmiiicon{scale=0.7}), SMX (\protect\smxicon{scale=0.7}) and FLIM (\protect\flimicon{scale=0.7}).}
	\label{fig_secomparison}
\end{figure}The factor $(M+1)$ in \eqref{eq_etaflim} emerges from the relaxed number of \glspl{LED}, by using the binomial theorem, which yields an enlarged transmit set. However, this advantage will diminish while $M$ increases. Note from \eqref{eq_etaflim} also that the relaxation of the number of active \glspl{LED} is essentially equivalent to extending the $M$-ary \gls{PAM} into $(M+1)$-ary constellation, $\bar{\mathcal{M}}= \{ I_0 \} \cup \mathcal{M}$. The \textit{zero} symbol is given by $I_0\in [0~ I_\text{L}]$. Therefore, unlike conventional systems, odd numbered $M$ values could be utilized in \gls{FLIM}. The spectral efficiency comparisons between \gls{FLIM} and the reference systems are provided in \figref{fig_secomparison} where it can be seen that the spectral efficiency advantage of \gls{FLIM} is still effective when $M=2$ and $M=4$. In practice, the \gls{PAM} modulation order is chosen to be within $2 \leq\ M \leq 8$ in order to achieve a feasible error performance.

Then, the transmit signal vector $\mathbf{s}_t$ is fed to the \gls{DAC} and \gls{E/O} before the transmission. The optical domain transmit vector passes through an $N_\textrm{t}\times N_\textrm{r}$ optical channel matrix $\mathbf{H}$. At the \gls{RX} side, the electrical domain received signal is obtained after the \gls{O/E} and \gls{ADC}. The baseband received signal vector $\mathbf{y}=[y_0,y_1,\cdots,y_{N_\textrm{r}}]^\textrm{T}\in\mathbb{R}_{N_\textrm{r}\times1}$ is given by $\mathbf{y}=\mathbf{r}_t+\mathbf{n}$, where $N_\textrm{r}\times 1$ optical received signal vector is given by $\mathbf{r}_t=\mathbf{H}\mathbf{s}_t=[r_{t,0},r_{t,1},\cdots,r_{t,N_\textrm{r}-1}]^\textrm{T}$. The $N_\textrm{r}\times 1$ \gls{AWGN} vector is denoted by $\mathbf{n}\in \mathbb{R}_{N_\textrm{r}\times1}$. The elements of $\mathbf{n}$ follow $\mathcal{N}(0,\sigma_{n}^2)$.  In this  paper, without loss of generality, the values of \gls{E/O}, \gls{O/E}, \gls{ADC} and \gls{DAC} coefficients are assumed to be one. Thus, the received electrical power expression is given as follows:
\begin{multline}
P_\textrm{elec}=\sum_{i=1}^{N_\textrm{r}}\textrm{E}\{r_{t,i}^2\}=\sum_{i=1}^{N_\textrm{r}}\sum_{j=1}^{N_\textrm{t}}H_{i,j}^2\textrm{E}\{s_{t,j}^2\}\\
+\sum_{i=1}^{N_\textrm{r}}\sum_{k=1}^{N_\textrm{t}}\sum\limits_{\substack{l=1 \\ k \neq l}}^{N_\textrm{t}} H_{i,k}H_{i,l}\textrm{E}\{s_{t,k}\}\textrm{E}\{s_{t,l}\}.
\label{eq_pelec}
\end{multline}
The \gls{PMF} for $i^\textrm{th}$ element of $\mathbf{s}_t$ is given by:
\begin{align}
p_{s_{t,i}}(x)=\frac{1-\nu_i}{M}+\nu_i \delta[x],
\label{eq_pmf}
\end{align}
\noindent where $\nu_i$ denotes the probability of a \textit{zero} occurring in $s_{t,i}$. It should be noted that the parameter $\nu_i$ changes depending on the adopted transmission technique. In \gls{SMX}, $\nu_i=0,\forall i$, whereas it is $\nu_i=1/N_\textrm{t},\forall i$ in the conventional \gls{SM}. For \gls{GSM}-II and \gls{FLIM}, the values of the $\nu_i$'s are a parameter of the chosen transmission subset. The electrical power of the $s_{t,i}$ in \eqref{eq_pelec} is calculated by using \eqref{eq_pmf} as
\begin{align}
\textrm{E}\{s_{t,i}^2\}= (1-\nu_i)\left( I_\textrm{L}^2+I_\textrm{L}(I_\textrm{U}-I_\textrm{L})+\frac{2M-1}{6(M-1)}(I_\textrm{U}-I_\textrm{L})^2 \right).
\label{eq_elecSNR}
\end{align}
\noindent Hence, the received electrical \gls{SNR} per bit becomes
\begin{align}
\frac{E_\textrm{b,elec}}{N_0}=\frac{P_\textrm{elec}}{\eta N_0}.
\label{eq_snr}
\end{align}
\noindent The parameter $N_0$ denotes the single sided noise power spectral density, $\sigma_n^2=N_0B$ where $B$ is the bandwidth of the effective noise process. In this paper, $B$ is assumed to be one without loss of generality. The spectral efficiency of the adopted modulation technique in terms of average bits per symbol is denoted by $\eta$.

Next, the \gls{ML} detector jointly estimates both the spatial and constellation symbols by using the received vector $\mathbf{y}$ as follows:
\begin{align}
\hat{\mathbf{s}}_t=\arg \max\limits_{\mathbf{s}_t \in \mathbb{S}} p(\mathbf{y} \vert \mathbf{H}\mathbf{s}_t)=\arg \min\limits_{\mathbf{s}_t \in \mathbb{S}} \norm{\mathbf{y}-\mathbf{H}\mathbf{s}_t}.
\label{eq_ML}
\end{align}
The feed-forward equalization block of the \gls{ML} detector is given by $\mathbf{F}=\mathbf{I}_{N_\textrm{r}}$. Lastly, the transmitted information bits are recovered by an inverse one-to-one mapping, $f_{\mathcal{M}}^{-1}$ as \mbox{$\hat{\mathbf{b}}_i=f_\mathcal{M}^{-1}\left(\hat{\mathbf{s}}_t\right)$}.
The \gls{BEP} for the optimal \gls{ML} detector is not straightforward to obtain due to the joint error probabilities of the transmission symbols. However, the \gls{BEP} could be upper bounded by the union bound as follows:
\begin{align}
P_\textrm{b}\leq \frac{1}{C\log_2 (C)}\sum_{i=1}^{C}\sum\limits_{\substack{j=1}}^{C}\textrm{d}_\textrm{H}\left( \mathbf{s}_i , \mathbf{s}_j \right)\textrm{P}( \mathbf{s}_i \rightarrow \mathbf{s}_j)
\label{eq_unionbound}
\end{align}
\noindent where the Hamming distance between the binary labels of $\mathbf{s}_i$ and $\mathbf{s}_j$ is denoted by $\textrm{d}_\textrm{H}\left( \mathbf{s}_i , \mathbf{s}_j \right)$. The \gls{PEP} between $\mathbf{s}_i$ and $\mathbf{s}_j$ is given by
\begin{align}
\textrm{P} (\mathbf{s}_i \rightarrow \mathbf{s}_j)=\textrm{Q}\left( \frac{\norm{\mathbf{H}(\mathbf{s}_i-\mathbf{s}_j)}}{2\sigma_n} \right).
\end{align}
It is worth noting from \eqref{eq_ML} that the \gls{ML} detector is optimal in terms of error probability as $\mathbf{s}_t$'s are equally likely. However, the size of the search space and the receiver complexity are proportional to the cardinality of the chosen set. It can be inferred from $\vert \mathbb{S}^\textrm{SMX} \vert$, $\vert \mathbb{S}^\textrm{GSM-II} \vert$ and $\vert \mathbb{S}^\textrm{FLIM} \vert$ that the search space is an exponential function of $N_\textrm{t}$. Therefore, in this paper, we are investigating an \gls{MMSE} based \gls{MIMO} detector which has a linear complexity with respect to $N_\textrm{t}$.
\subsection{Low Complexity MMSE Detector}
Unlike the optimal \gls{ML} detector, which determines both spatial and constellation symbols jointly, the linear \gls{MMSE} detector treats the elements of the transmission vector individually. Firstly, the coupling effect of the parallel \gls{MIMO} channel is reversed by a feed-forward \gls{MMSE} filter $\mathbf{F}$ as
\begin{align}
\mathbf{F} = \mathbf{R_s} \mathbf{H}^\text{T} \left( \mathbf{H} \mathbf{R_s} \mathbf{H}^\text{T} + \mathbf{R_w} \right)^{-1},
\label{eq_MMSE}
\end{align}
\noindent where $\mathbf{R_s}\triangleq\text{E}\{ \mathbf{s}_t \mathbf{s}^\text{T}_t  \}$ is the autocorrelation matrix averaged over all the possibilities of $\mathbf{s}_t$. The autocorrelation matrix of the coloured noise after filtering, $\hat{\mathbf{y}}=\mathbf{F}\mathbf{y}$, is given by $\mathbf{R_w} = \textrm{E}\{ \mathbf{w}\mathbf{w}^\textrm{T} \}=\sigma_{w}^2 \mathbf{I}_{N_\textrm{r}}$. Secondly, the transmitted symbols are detected by an element-wise detector as follows:
\begin{align}
\hat{s}_i=\arg \max\limits_{\substack{0\leq k \leq M \\ s_k \in \mathcal{M}}} p(\hat{y}_i \vert s_k)=\arg \min\limits_{\substack{0\leq k \leq M \\ s_k \in \mathcal{M}}} \vert \hat{y}_i-s_k \vert.
\label{eq_MLMMSE}
\end{align}
\noindent It should be noted from \eqref{eq_MLMMSE} that the number of active \glspl{LED} is relaxed in \gls{FLIM} as given by \eqref{eq_setFLIM}. Thus, the \gls{RX} has no information about the number of active \glspl{LED} in the transmitted symbols, whereas this is not the case in \gls{SMX} and \gls{SM}-based systems. Therefore, $I_0$ can be considered as one of the constellation symbols in \gls{FLIM} as mentioned previously. By using \eqref{eq_pam}, the $M$-ary \gls{PAM} constellation set is extended to $\bar{\mathcal{M}}=\{0\}\cup \mathcal{M}$ for $k\in \{0,1,\cdots,M\}$. The singleton $\{0\}$ corresponds to the $(M+1)^\textrm{th}$ element of the $\bar{\mathcal{M}}$.
\section{Computer Simulations}\label{sec:simulationsandresults}
In this section, computer simulation results for $4\times 4$ systems are presented to first investigate the effect of an \gls{ADR} on the \gls{CN} distribution. Then, the \gls{BEP} performance comparisons between \gls{FLIM}, \mbox{\gls{GSM}-II} and \gls{SMX} when they employ \gls{ADR} are obtained by the use of Monte Carlo simulations. The optical attocell structure considered in the simulations is depicted in \figref{fig_ledpdmodel}(a). The location of the mobile \gls{UE} within the attocell is defined by polar coordinates, $(R_{\textrm{UE}}, \omega_{\textrm{UE}})$. The radius and polar angle parameters are denoted by $R_{\textrm{UE}}$ and $\omega_{\textrm{UE}}$, respectively where $0 \leq R_{\textrm{UE}} \leq R_{\textrm{cell}}$ and $0\leq \omega_\textrm{UE} \leq 2\pi$. The cell radius, as depicted in \figref{fig_ledpdmodel}, is defined as the region where the maximum optical power received from an \gls{LED} is halved. Thus, $R_{\text{cell}}=\rho_{\text{cell}}\left( H_{\text{lum}}-H_{\text{UE}} \right)$ where $\rho_{\text{cell}}=\sqrt{4^{1/(m+3)}-1}$. The parameters, $H_{\textrm{lum}}$ and $H_{\textrm{UE}}$, denote the height of the luminaire and the distance of the \gls{UE} from the floor, respectively. Their values are adopted from \cite{8422436}. Moreover, the \glspl{LED} are chosen as Lambertian transmitters with $\Phi_{1/2}=60^\circ$. Similarly, \glspl{PD} are assumed to have a \gls{FoV} of $85^\circ$. The entire set of simulation parameters and their values are given in Table~\ref{table_simpar}.

\begin{figure}[!t]
	\centering
	\begin{subfigure}[b]{.5\columnwidth}
		\centering
		\includegraphics[width=0.85\columnwidth,draft=false]{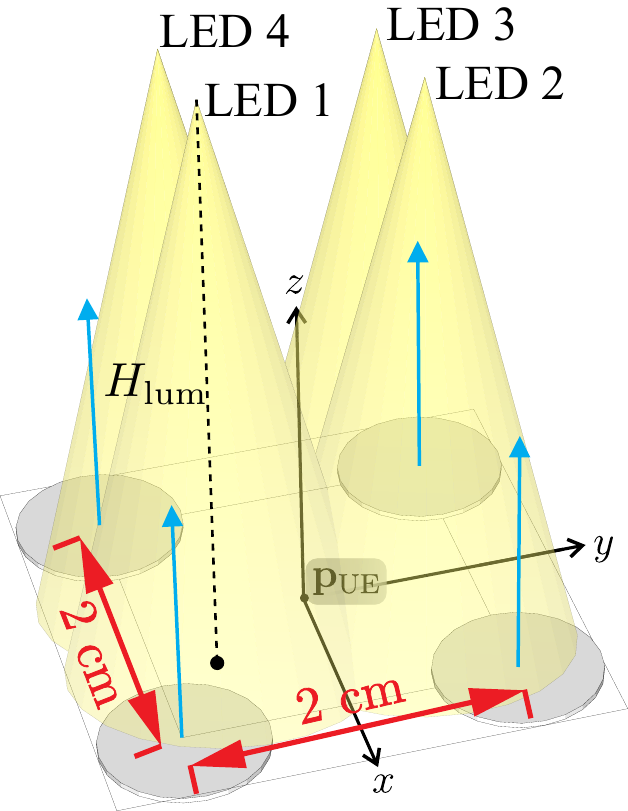}
		\caption{Square configuration}
	\end{subfigure}~
	\begin{subfigure}[b]{.5\columnwidth}
		\centering
		\includegraphics[width=0.85\columnwidth,draft=false]{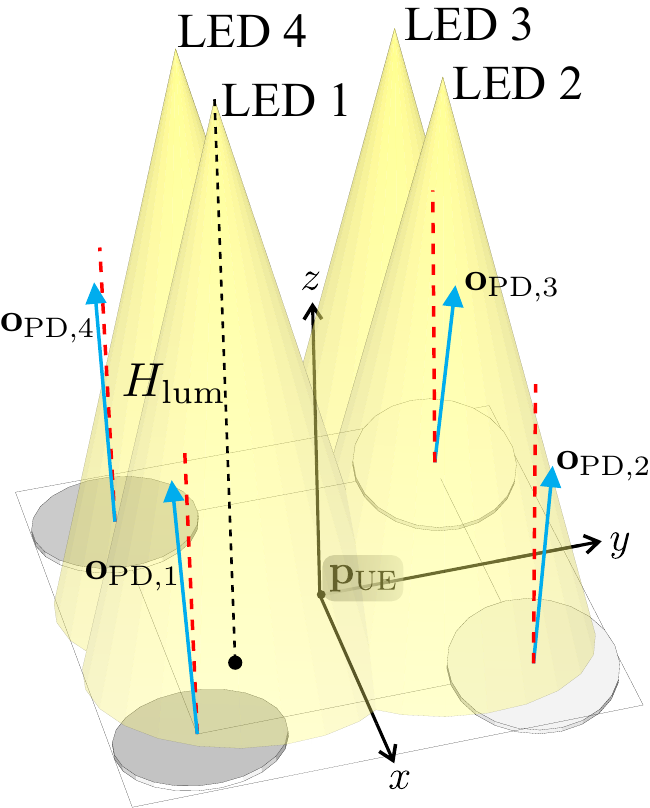}
		\caption{Square-perturbed configuration}
	\end{subfigure}
	\caption{Structure of the (a) typical receiver (a) and (b) ADR.}
	\label{fig_pdconfig}
\end{figure}

In this work, two \gls{PD} configurations of the receivers; square and square-perturbed, which are depicted in \figref{fig_pdconfig}(a) and \figref{fig_pdconfig}(b), respectively are considered. Accordingly, in the square receiver, each \gls{LED} is aligned with a \gls{PD} where the separation between the \glspl{PD} is $2$ cm. In the perturbed square receiver, the \gls{PD} formation stays the same however, random polar and azimuth angles are introduced to each \gls{PD} in order to break the potential rank deficiency. The polar and azimuth angles of the perturbed receiver are taken as $\left\{\{\gamma_j,\beta_j\}_1^{N_\text{r}}\right\} = \left\{\{-5,6\}, \{-8,1\}, \{-10,2\}, \{15,1\} \right\}$ degrees. It is worth noting that any arbitrary selection of the polar and azimuth angles will break the symmetry in the channel matrix. A deeper investigation on the tilt angles and their optimal values will be carried out in future work.
\begin{table}[!t]
	\centering
	\caption{Simulation Parameters}
	\label{table_simpar}
	\resizebox{1\columnwidth}{!}{
		\renewcommand{\arraystretch}{1.35} 
		\begin{tabular}{|l|l|l|}
			\hline
			\textbf{Parameter} 			 & \textbf{Description}                                    		  & \textbf{Value} \\ \hline\hline
			$R_{\text{cell}}$            & Radius of the optical attocell.                                & $100$ cm     \\ \hline
			$L$            			 & Separation between two LEDs.									  & $2$ cm              \\ \hline
			$H_{\text{lum}}$             & Height of the luminaire (optical AP).                                        & $300$ cm              \\ \hline
			$H_{\text{human}}$           & Average height of the human.                                   & $180$ cm              \\ \hline
			$H_{\text{UE}}$          & Average height of the mobile device.                           & $144$ cm              \\ \hline
			$A_{\text{PD}}$              & Area of the PD.                                                & $1$ $\text{cm}^2$              \\ \hline
			$\Phi_{1/2}$                 & Semi-angle of half power of the LED.                           & $60^\circ$              \\ \hline
			$\kappa_{\text{1/2}}$     & FoV of the PD.                                       & $85^\circ$              \\ \hline
			$\omega_\textrm{UE}$               		 & Polar angle of the mobile user location.                        & $[0~2\pi]$ rad              \\ \hline
			$R_{\textrm{UE}}$               			 & Radius of the mobile user location.                       & $[0~100]$ cm              \\ \hline
			$I_{\textrm{U}}$               			 & The upper limit for the $I_\textrm{f}$.                       & $800$ mA \cite{creexlampxml}             \\ \hline
			$I_{\textrm{L}}$               			 & The lower limit for the $I_\textrm{f}$.                       & $500$ mA \cite{creexlampxml}              \\ \hline
		\end{tabular}
	}
\end{table}

\begin{figure}
	\centering
	\begin{subfigure}[b]{0.5\columnwidth}
		\centering
		\includegraphics[width=1\columnwidth,draft=false]{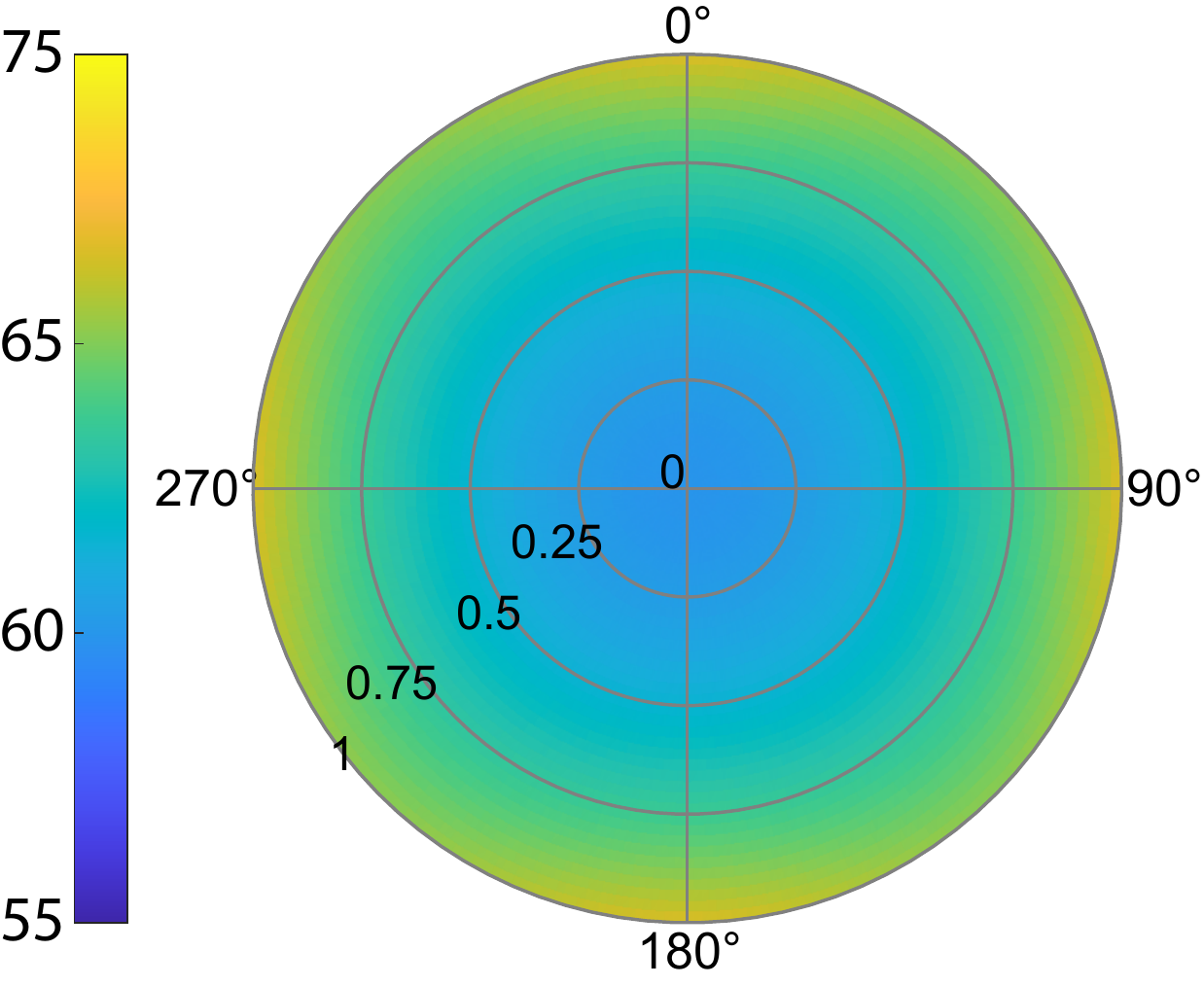}
		\caption{Square configuration}
	\end{subfigure}~
	\begin{subfigure}[b]{0.5\columnwidth}
		\centering
		\includegraphics[width=1\columnwidth,draft=false]{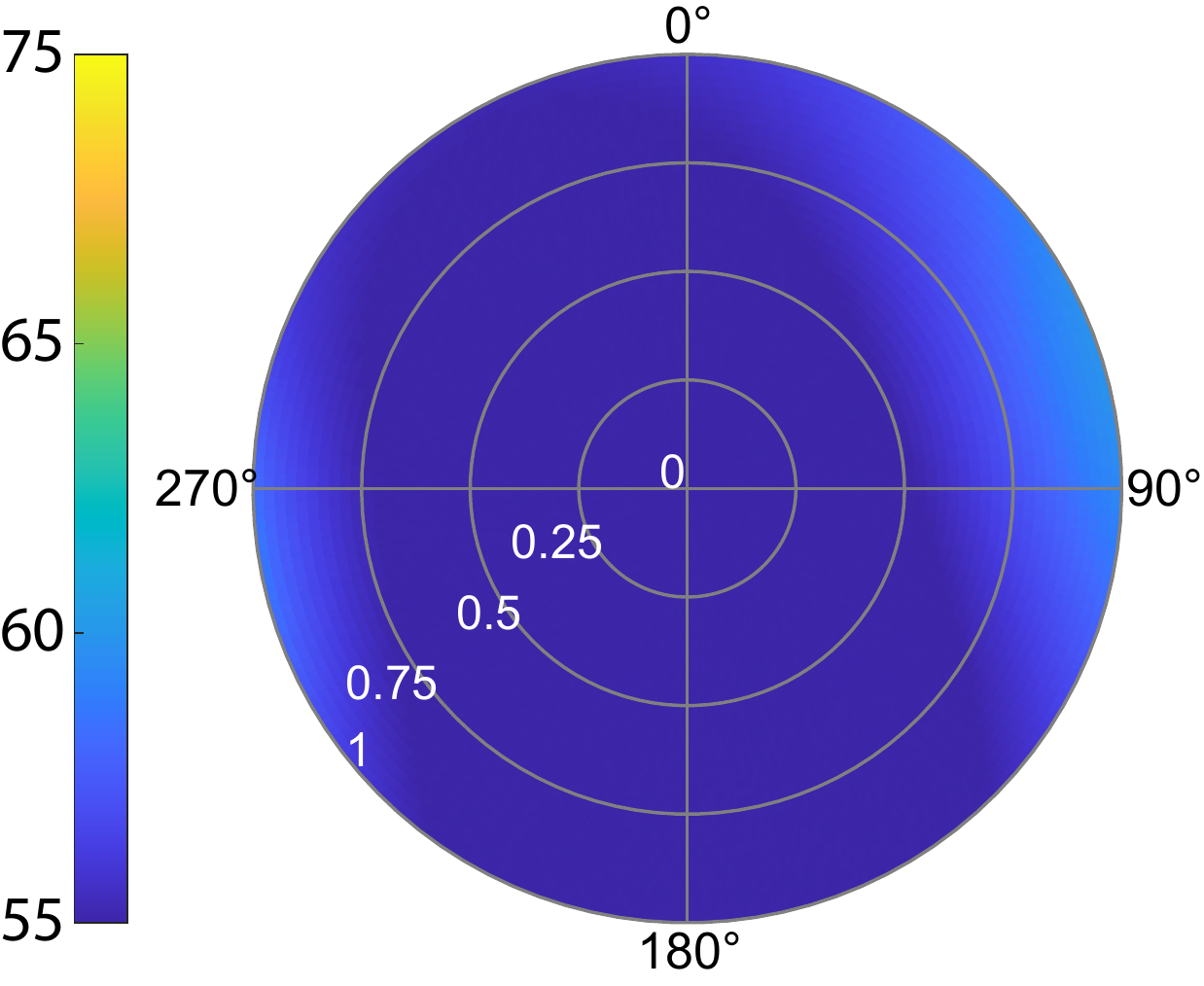}
		\caption{Square-perturbed configuration}
	\end{subfigure}
	\caption{CN distribution plots (in dB) for adopted receiver configurations.}
	\label{fig_condnum}
\end{figure}

The performance of the suboptimal \gls{MMSE} detector is closely related to the \gls{CN}, as stated in the previous section. Therefore, the channel matrix \gls{CN} distributions for a mobile \gls{UE} with the adopted \gls{PD} structures is depicted in \figref{fig_condnum}, by using polar plots. The values of the \gls{CN} are given for any location of the mobile \gls{UE} within an attocell. The resolution of the $R_{\textrm{UE}}$ and $\omega_{\textrm{UE}}$ are taken as $2.5$ centimetres and $1$ degree, respectively. The \gls{CN} distribution of the square receiver is given in \figref{fig_condnum}(a). As can be seen from the figure, the \gls{CN} of the square receiver is approximately $8$ dB smaller in the cell centre compared to the cell edges. It can be inferred from \eqref{eq_dcchannelgain} that the \gls{MIMO} channel matrix elements get smaller at the cell edges due to the inverse square distance law. Thus, the linear dependency between the rows and/or columns increases, which effectively yields a high \gls{CN}. The average \gls{CN} achieved by the square receiver becomes $62.14$ dB with a standard deviation of $1.89$ dB.

In \figref{fig_condnum}(b), the \gls{CN} distribution of the square-perturbed receiver is depicted for the same scale. As seen from the figure, the \gls{CN} is dramatically reduced as the \gls{MIMO} channel elements are differentiated with the aid of random perturbation. Moreover, the distribution of the \gls{CN} is almost flat, except for some small areas near the cell edges. Again, due to the inverse square law and random direction of the \glspl{PD}, relatively high \gls{CN} regions are observed in the north-east and south-west edges of the polar plot. The square-perturbed receiver structure achieves an average \gls{CN} of $54.11$ dB with the standard deviation of $1.52$ dB. Hence, the random perturbation of the \glspl{PD} reduces the average \gls{CN} and standard deviation by around $8$ and $0.37$ dB, respectively. Consequently, in \gls{ABEP} simulations, both proposed \gls{FLIM} and the reference systems are assumed to be utilizing the square-perturbed structure at the receiving \gls{UE}.

\begin{figure}[!t]
	\centering
	\begin{subfigure}[b]{.5\columnwidth}
		\centering
		\includegraphics[width=1\columnwidth,draft=false]{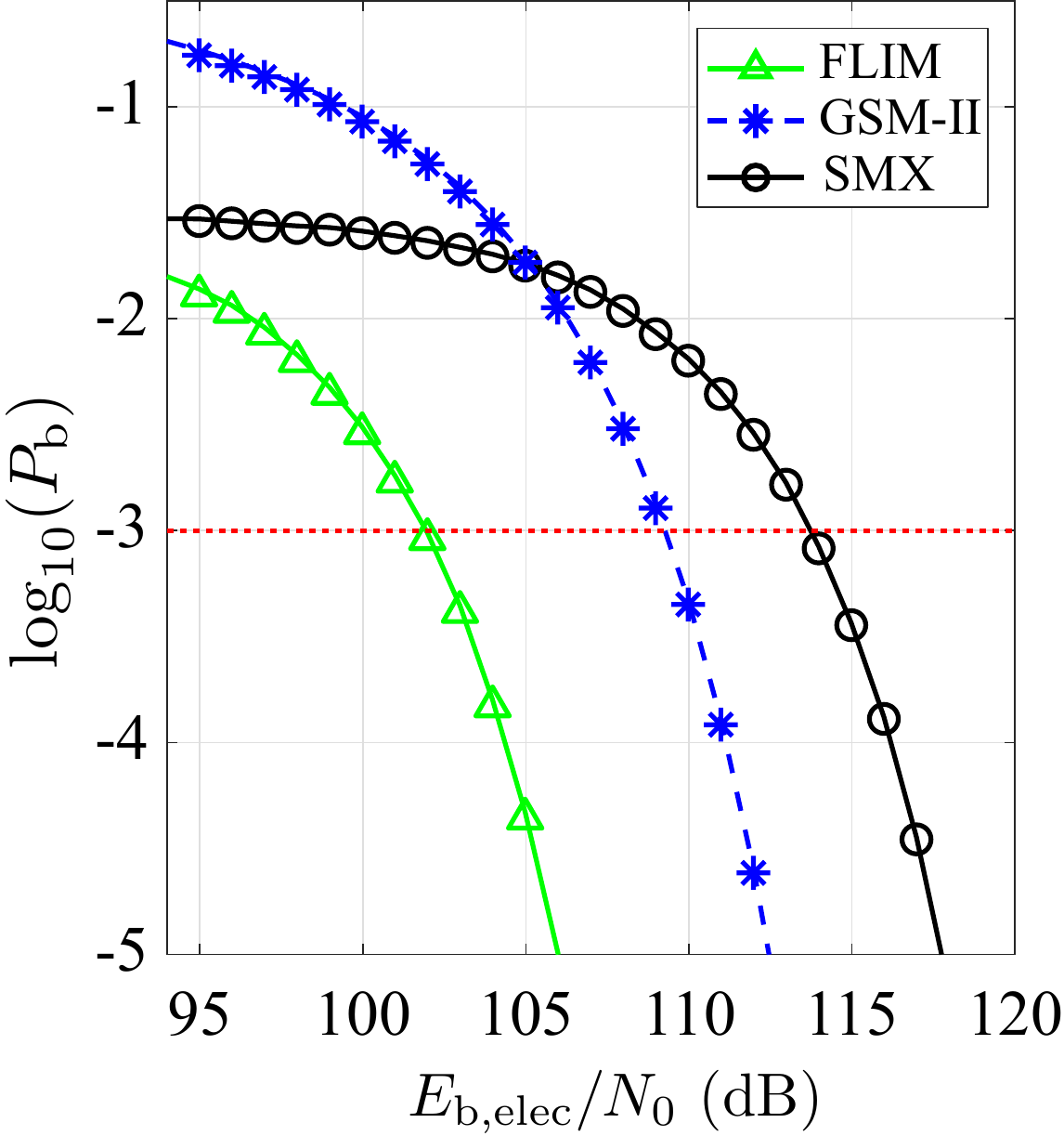}
		\caption{$\eta=4$ bpcu and cell centre}
	\end{subfigure}~
	\begin{subfigure}[b]{.5\columnwidth}
		\centering
		\includegraphics[width=1\columnwidth,draft=false]{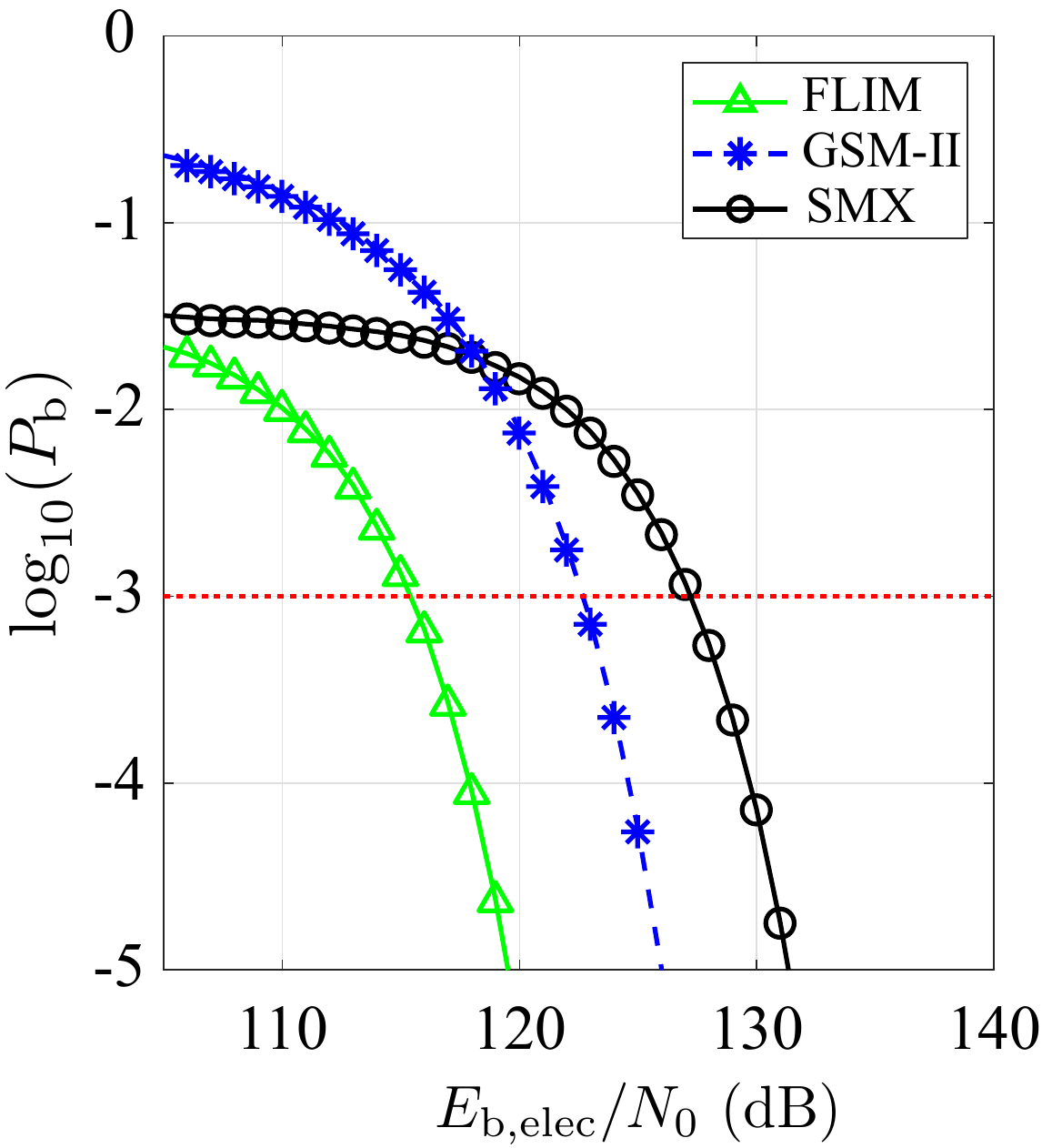}
		\caption{$\eta=4$ bpcu and cell edge}
	\end{subfigure}\\
	\begin{subfigure}[b]{.5\columnwidth}
		\centering
		\includegraphics[width=1\columnwidth,draft=false]{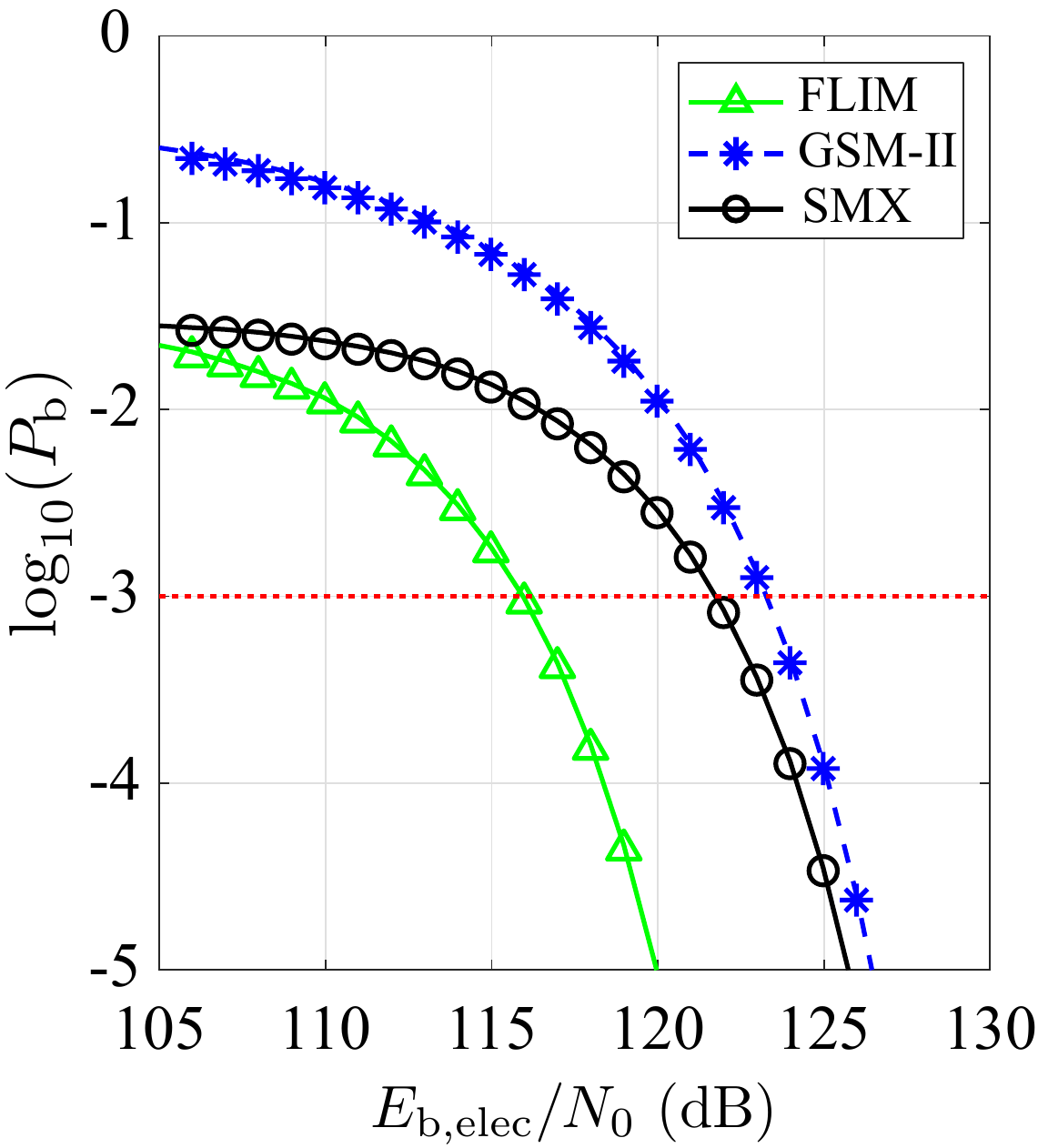}
		\caption{$\eta=8$ bpcu and cell centre}
	\end{subfigure}~
	\begin{subfigure}[b]{.5\columnwidth}
		\centering
		\includegraphics[width=1\columnwidth,draft=false]{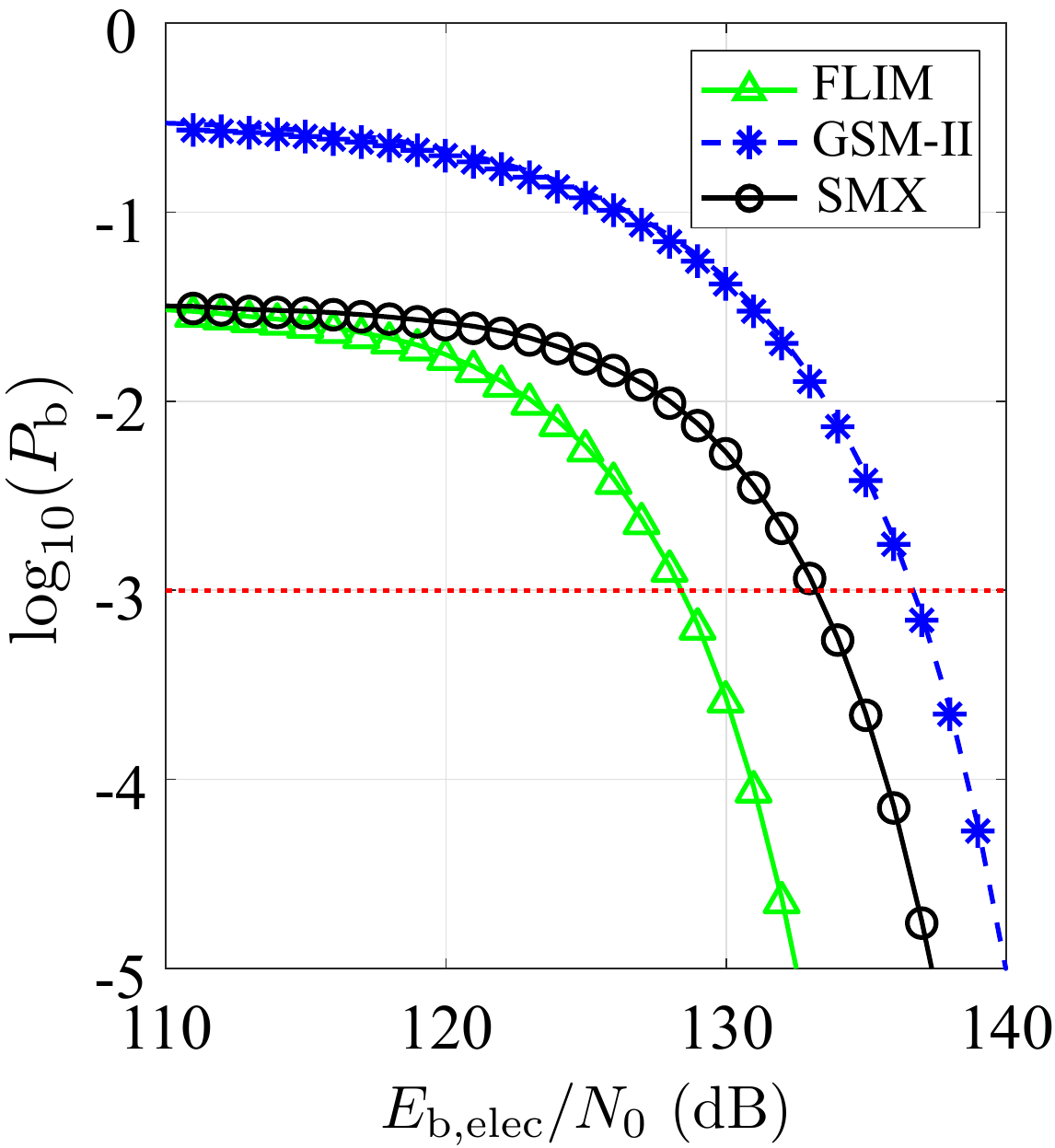}
		\caption{$\eta=8$ bpcu and cell edge}
	\end{subfigure}
	\caption{ABEP vs. transmit $E_{\text{b,elec}}/N_0$ comparison for GSM-II, SMX and FLIM systems at: (a) ($R_\textrm{UE}=0,~\omega_\textrm{UE}=0^\circ$) and $\eta=4$ bpcu, (b) ($R_\textrm{UE}=1,~\omega_\textrm{UE}=90^\circ$) and $\eta=4$ bpcu, (c) ($R_\textrm{UE}=0,~\omega_\textrm{UE}=0^\circ$) and $\eta=8$ bpcu and (d) ($R_\textrm{UE}=1,~\omega_\textrm{UE}=90^\circ$) and $\eta=8$ bpcu.}
	\label{fig_bercomparison}
\end{figure}

In \figref{fig_bercomparison}, the \gls{ABEP} comparisons for the \gls{FLIM}, \gls{GSM}-II and \gls{SMX} are given with respect to the transmit electrical \gls{SNR} per bit. Accordingly, all three systems employ low-complexity linear \gls{MMSE} for symbol detection and a square-perturbed \gls{PD} structure for enhanced detector performance. The plots are given for two spectral efficiency values; low ($\eta=4$ bpcu), in Figs. \ref{fig_bercomparison}(a), (b), and medium/high ($\eta=8$ bpcu), in Figs. \ref{fig_bercomparison}(c), (d). Furthermore, the impact of the \gls{CN} on the \gls{ABEP} performance is investigated for cell centre ($R_\textrm{UE}=0,~\omega_\textrm{UE}=0^\circ$) and edge ($R_\textrm{UE}=1,~\omega_\textrm{UE}=90^\circ$) locations of the \gls{UE}, depicted in Figs. \ref{fig_bercomparison}(a), (c) and Figs. \ref{fig_bercomparison}(b), (d), respectively. It is important to note that the \gls{ABEP} curves are obtained by running Monte-Carlo simulations for $10^7$ transmit symbol realizations. In order to fairly compare the \glspl{BEP} of multiple systems under different channel conditions, the transmit electrical \gls{SNR} expression obtained by \eqref{eq_elecSNR} must be used. Since the magnitude of the elements in the optical \gls{MIMO} channel is in the order of $10^{-4}$ and $10^{-5}$ for the cell centre and edges, respectively, the electrical path loss at the receiver becomes $-80$ and $-100$ dB. Both \gls{FLIM} and \gls{GSM}-II techniques have some degree of flexibility in the transmission signal set design. Thus, in our simulations, a suboptimal signal set design and bit labelling algorithm is adopted for \gls{FLIM} in order to reduce the system complexity. Accordingly, the suboptimal design matches the minimum Euclidean distance elements with the largest Hamming distances. Whereas \gls{GSM}-II takes the advantage of optimal signal set design and bit labelling which is subject to minimize \eqref{eq_unionbound}. Since the entire transmission symbols set is used in \gls{SMX}, only the Gray bit labelling is adopted. Lastly, the number of active \glspl{LED} are fixed to the number $N_\textrm{a}=N_\textrm{t}/2$ in \gls{GSM}-II as this yields the maximum spectral efficiency.

In Figs. \ref{fig_bercomparison}(a) and (b), the \gls{ABEP} vs. transmit $E_{\text{b,elec}}/N_0$ performance of \gls{FLIM} is compared with \gls{GSM}-II and \gls{SMX} for $\eta=4$ bpcu when the mobile \gls{UE} is located at the cell centre and edge, respectively. As can be seen from both figures, \gls{FLIM} outperforms \gls{GSM}-II and \gls{SMX} approximately $7$ and $11$ dB, respectively in the high \gls{SNR} regime. The performance gain in \gls{FLIM}, compared to the reference systems, comes from the smaller modulation order requirement to achieve the target bit rate. Accordingly, the $M$ values in order to achieve $\eta=4$ bpcu with \gls{FLIM}, \gls{GSM}-II and \gls{SMX} systems become, $M_{\text{SMX}}=2$, \mbox{$M_{\text{GSM-II}}=2$} and $M_{\text{FLIM}}=1$. Unlike \gls{GSM}-II and \gls{SMX}, \gls{FLIM} is able to retain high multiplexing gains without sacrificing the energy efficiency benefits of the \gls{LED} index domain utilization. The location of the mobile \gls{UE} and related channel matrix coefficients are other important factors for the system performance. The difference between the curves when the \gls{UE} is located at the cell centre and edge is around $13$ dB in the high \gls{SNR} regime for $\eta=4$ bpcu. The reason for this is two-fold; the difference in (i) \gls{CN} and (ii) element-wise magnitudes of the \gls{MIMO} channel matrices. Specifically, \glspl{CN} at the cell centre and edge are $52.76$ and $59.41$ dB, respectively. Furthermore, the magnitudes of the channel matrix elements are approximately $10$ times larger at the cell centre compared to the edge.

In Figs. \ref{fig_bercomparison}(c) and (d), the plots are given for $\eta=8$ bpcu, where $M_{\text{SMX}}=4$, $M_{\text{GSM-II}}=8$ and $M_{\text{FLIM}}=3$. Similarly, \gls{FLIM} outperforms \gls{GSM}-II and \gls{SMX} by approximately $7$ and $6$ dB in Fig. \ref{fig_bercomparison}(c) and $8$ and $4$ dB in in Fig. \ref{fig_bercomparison}(d), respectively. Again, \gls{FLIM} takes advantage of its extended transmission symbols set with a smaller modulation order. It is important to note that the \gls{SMX} outperforms \gls{GSM}-II both in the cell centre and the edge for $\eta=8$ bpcu. This can also be explained by the higher modulation size requirement of the \gls{GSM}-II to achieve the target bit rate. In parallel, with the low spectral efficiency application, the difference between all the systems when the \gls{UE} is located at the cell centre and edge also becomes $13$ dB in the high \gls{SNR} regime. Lastly, the curves for the \gls{FLIM}, \gls{GSM}-II and \gls{SMX} at the cell centre are shifted to the right-hand side by $14$, $14$ and $8$ dB, respectively when the target bit rate is doubled. For the cell edge, the values become $13$, $14$ and $6$ dB, respectively. This is due to the reduced minimum Euclidean distance in the transmission symbols set that is imposed by the increase in $M$.
\section{Conclusions}
In this paper, a transmission technique for \gls{MIMO}-\gls{OWC}, introduced as \gls{FLIM}, has been presented. Unlike conventional systems, the proposed method takes advantage of the relaxed number of active \glspl{LED} per time instance. Thus, the transmission symbol set is greatly extended in \gls{FLIM}. The additional symbols can be used to enhance the error performance and/or power efficiency by subset selection. Moreover, a perturbed receiver structure has also been proposed in order to cope with the rank deficiency which paves the way for low-complexity \gls{MMSE} detection. Computer simulations results have shown that \gls{FLIM} achieves a better error performance compared to \gls{GSM}-II and \gls{SMX}. Consequently, the enhanced multiplexing gain and power efficiency makes \gls{FLIM} a highly suitable candidate for next generation \gls{MIMO}-\gls{OWC} applications. The complexity/time analysis of the proposed technique and the optimal tilt angle derivations are reserved for future work.


%




\ifCLASSOPTIONcaptionsoff
  \newpage
\fi



%
%
%
\newpage
\bibliography{IEEEabrv,kitap2018}
\bibliographystyle{IEEEtran}

%








\end{document}